\documentstyle[12pt]{article}

\makeatletter
\def\section{\@startsection {section}{1}{\z@}{-3.5ex plus -1ex minus
 -.2ex}{2.3ex plus .2ex}{\large\bf}}
\def\subsection{\@startsection{subsection}{2}{\z@}{-3.25ex plus -1ex minus
 -.2ex}{1.5ex plus .2ex}{\normalsize\bf}}
\makeatother

\makeatletter
\@addtoreset{equation}{section}

\makeatother

\newcommand{\nc}{\newcommand}
\newcommand{\rnc}{\renewcommand}

\nc{\be}{\begin{equation}}
\nc{\ee}{\end{equation}}
\nc{\bea}{\begin{eqnarray}}
\nc{\eea}{\end{eqnarray}}

\rnc{\a}{\alpha}
\rnc{\b}{\beta}
\rnc{\gg}{\gamma}
\rnc{\d}{\delta}
\nc{\e}{\eta}
\nc{\eb}{\bar{\eta}}
\nc{\ep}{\epsilon}
\nc{\f}{\phi}
\nc{\fb}{\bar{\phi}}
\nc{\vf}{\varphi}
\nc{\p}{\psi}
\rnc{\pb}{\bar{\psi}}
\rnc{\c}{\chi}
\nc{\cb}{\bar{\c}}
\nc{\la}{\lambda}
\nc{\m}{\mu}
\nc{\n}{\nu}
\rnc{\o}{\omega}
\nc{\Om}{\Omega}
\rnc{\t}{\theta}
\nc{\eps}{\epsilon}

\rnc{\S}{\Sigma}
\nc{\Sa}{\S\times\{0\}}
\nc{\Sb}{\S\times\{1\}}
\nc{\SI}{\S\times I}
\nc{\SS}{\S\times S^{1}}
\nc{\M}{{\cal M}}

\nc{\trac}[2]{{\textstyle\frac{#1}{#2}}}

\nc{\ex}[1]{{\rm e}^{\,\textstyle#1}}

\nc{\mat}[4]{\left(\begin{array}{cc}#1&#2\\#3&#4\end{array}\right)}

\nc{\som}[9]{\left(\begin{array}{ccc}#1&#2&#3\\#4&#5&#6\\#7&#8&#9%
\end{array}\right)}

\def\Tr{\mathop{\rm Tr}\nolimits}
\nc{\ra}{\rightarrow}
\nc{\ot}{\otimes}
\rnc{\ss}{\subset}
\nc{\nul}{\noindent\underline}
\nc{\subs}[1]{{\vspace*{0.5cm}}%
{\noindent\underline{#1}}{\addcontentsline{toc}{subsection}{#1}}%
{\vspace*{0.3cm}}}

\rnc{\lg}{{\bf g}}
\nc{\lt}{{\bf t}}
\nc{\lk}{{\bf k}}
\nc{\bft}{{\bf t}}
\nc{\bfk}{{\bf k}}
\nc{\bfg}{{\bf g}}
\nc{\del}{\partial}
\nc{\dz}{\del_{z}}
\nc{\dzb}{\del_{\bar{z}}}
\nc{\az}{A_{z}}
\nc{\azb}{A_{\bar{z}}}
\nc{\bz}{B_{z}}
\nc{\bzb}{B_{\bar{z}}}
\nc{\ba}{{\bf A}}
\nc{\bb}{{\bf B}}
\nc{\g}{g^{-1}}
\nc{\dw}{\Delta_{W}}
\nc{\Det}{{\rm Det}\,}
\nc{\Ad}{{\rm Ad}}
\nc{\bG}{{\bf G}}
\nc{\bT}{{\bf T}}
\nc{\bK}{{\bf H}}
\nc{\bH}{{\bf H}}
\nc{\bP}{{\bf P}}
\nc{\gt}{\bG/\bT}
\nc{\map}[2]{{\rm Map}(#1,#2)}
\nc{\mg}{\map{M}{\bG}}
\nc{\mgr}{\map{M}{\bG_{r}}}
\nc{\mt}{\map{M}{\bT}}
\nc{\mtr}{\map{M}{\bT_{r}}}
\nc{\sg}{\map{\Sigma}{\bG}}
\nc{\sgr}{\map{\Sigma}{\bG_{r}}}
\nc{\st}{\map{\Sigma}{\bT}}
\nc{\str}{\map{\Sigma}{\bT_{r}}}
\nc{\mgt}{\map{M}{\gt}}
\nc{\sgt}{\map{\Sigma}{\gt}}
\nc{\ug}{\map{U}{\bG}}
\nc{\ugr}{\map{U}{\bG_{r}}}
\nc{\ut}{\map{U}{\bT}}
\nc{\utr}{\map{U}{\bT_{r}}}
\nc{\uag}{\map{\ua}{\bG}}
\nc{\uagr}{\map{\ua}{\bG_{r}}}
\nc{\uat}{\map{\ua}{\bT}}
\nc{\uatr}{\map{\ua}{\bT_{r}}}
\nc{\uabg}{\map{\uab}{\bG}}
\nc{\uabgr}{\map{\uab}{\bG_{r}}}
\nc{\uabt}{\map{\uab}{\bT}}
\nc{\uabtr}{\map{\uab}{\bT_{r}}}
\nc{\unA}{\underline{A}}
\nc{\C}{{\cal A}/{\cal G}}
\nc{\A}[1]{{\cal A}^{#1}/{\cal G}^{#1}}
\nc{\dx}{\dot{x}}
\rnc{\O}[2]{\Omega^{#1}({#2},\lg)}
\nc{\wif}{Weyl integral formula}
\nc{\CS}{Chern-Simons theory}
\nc{\tg}{\tilde{g}}
\nc{\tti}{\tilde{t}}
\nc{\th}{\tilde{h}}

\nc{\ga}{g_{\a}}
\nc{\gb}{g_{\b}}
\nc{\gc}{g_{\gg}}
\nc{\ha}{h_{\a}}
\nc{\hb}{h_{\b}}
\nc{\hc}{h_{\gg}}
\nc{\ka}{k_{\a}}
\nc{\kb}{k_{\b}}
\nc{\kc}{k_{\gg}}
\nc{\ta}{t_{\a}}
\nc{\tb}{t_{\b}}
\nc{\tc}{t_{\gg}}
\nc{\gab}{g_{\a\b}}
\nc{\gac}{g_{\a\gg}}
\nc{\gbc}{g_{\b\gg}}
\nc{\hab}{h_{\a\b}}
\nc{\hac}{h_{\a\gg}}
\nc{\hbc}{h_{\b\gg}}
\nc{\kab}{k_{\a\b}}
\nc{\kac}{k_{\a\gg}}
\nc{\kbc}{k_{\b\gg}}
\nc{\tab}{t_{\a\b}}
\nc{\tac}{t_{\a\gg}}
\nc{\tbc}{t_{\b\gg}}
\nc{\ua}{U_{\a}}
\nc{\ub}{U_{\b}}
\nc{\uc}{U_{\gg}}
\nc{\uab}{U_{\a}\cap U_{\b}}
\nc{\uac}{U_{\a}\cap U_{\gg}}
\nc{\ubc}{U_{\b}\cap U_{\gg}}

\makeatletter

\def\diagram{\leftwidth=\z@ \rightwidth=\z@ \topheight=\z@
\botheight=\z@ \setbox\@picbox\hbox\bgroup}

\def\enddiagram{\egroup\wd\@picbox\rightwidth\unitlength
\ht\@picbox\topheight\unitlength \dp\@picbox\botheight\unitlength
\hskip\leftwidth\unitlength\box\@picbox}

\def\bfig{\begin{diagram}}
\def\efig{\end{diagram}}
\newcount\wideness \newcount\leftwidth \newcount\rightwidth
\newcount\highness \newcount\topheight \newcount\botheight

\def\ratchet#1#2{\ifnum#1<#2 \global #1=#2 \fi}

\def\putbox(#1,#2)#3{%
\horsize{\wideness}{#3} \divide\wideness by 2
{\advance\wideness by #1 \ratchet{\rightwidth}{\wideness}}
{\advance\wideness by -#1 \ratchet{\leftwidth}{\wideness}}
\vertsize{\highness}{#3} \divide\highness by 2
{\advance\highness by #2 \ratchet{\topheight}{\highness}}
{\advance\highness by -#2 \ratchet{\botheight}{\highness}}
\put(#1,#2){\makebox(0,0){$#3$}}}

\def\putlbox(#1,#2)#3{%
\horsize{\wideness}{#3}
{\advance\wideness by #1 \ratchet{\rightwidth}{\wideness}}
{\ratchet{\leftwidth}{-#1}}
\vertsize{\highness}{#3} \divide\highness by 2
{\advance\highness by #2 \ratchet{\topheight}{\highness}}
{\advance\highness by -#2 \ratchet{\botheight}{\highness}}
\put(#1,#2){\makebox(0,0)[l]{$#3$}}}

\def\putrbox(#1,#2)#3{%
\horsize{\wideness}{#3}
{\ratchet{\rightwidth}{#1}}
{\advance\wideness by -#1 \ratchet{\leftwidth}{\wideness}}
\vertsize{\highness}{#3} \divide\highness by 2
{\advance\highness by #2 \ratchet{\topheight}{\highness}}
{\advance\highness by -#2 \ratchet{\botheight}{\highness}}
\put(#1,#2){\makebox(0,0)[r]{$#3$}}}

\def\adjust[#1]{} 

\newcount \coefa
\newcount \coefb
\newcount \coefc
\newcount\tempcounta
\newcount\tempcountb
\newcount\tempcountc
\newcount\tempcountd
\newcount\xext
\newcount\yext
\newcount\xoff
\newcount\yoff
\newcount\gap%
\newcount\arrowtypea
\newcount\arrowtypeb
\newcount\arrowtypec
\newcount\arrowtyped
\newcount\arrowtypee
\newcount\height
\newcount\width
\newcount\xpos
\newcount\ypos
\newcount\run
\newcount\rise
\newcount\arrowlength
\newcount\halflength
\newcount\arrowtype
\newdimen\tempdimen
\newdimen\xlen
\newdimen\ylen
\newsavebox{\tempboxa}%
\newsavebox{\tempboxb}%
\newsavebox{\tempboxc}%

\newdimen\w@dth

\def\setw@dth#1#2{\setbox\z@\hbox{$#1$}\w@dth=\wd\z@
\setbox\@ne\hbox{$#2$}\ifnum\w@dth<\wd\@ne \w@dth=\wd\@ne \fi
\advance\w@dth by 1.2em}

\def\t@^#1_#2{\def\n@one{#1}\def\n@two{#2}\mathrel{\setw@dth{#1}{#2}
\mathop{\hbox to \w@dth{\rightarrowfill}}\limits
\ifx\n@one\empty\else ^{\box\z@}\fi
\ifx\n@two\empty\else _{\box\@ne}\fi}}
\def\t@@^#1{\@ifnextchar_ {\t@^{#1}}{\t@^{#1}_{}}}
\def\to{\@ifnextchar^ {\t@@}{\t@@^{}}}

\def\t@left^#1_#2{\def\n@one{#1}\def\n@two{#2}\mathrel{\setw@dth{#1}{#2}
\mathop{\hbox to \w@dth{\leftarrowfill}}\limits
\ifx\n@one\empty\else ^{\box\z@}\fi
\ifx\n@two\empty\else _{\box\@ne}\fi}}
\def\t@@left^#1{\@ifnextchar_ {\t@left^{#1}}{\t@left^{#1}_{}}}
\def\toleft{\@ifnextchar^ {\t@@left}{\t@@left^{}}}

\def\two@^#1_#2{\def\n@one{#1}\def\n@two{#2}\mathrel{\setw@dth{#1}{#2}
\mathop{\vcenter{\hbox to \w@dth{\rightarrowfill}\kern-1.7ex
                 \hbox to \w@dth{\rightarrowfill}}%
       }\limits
\ifx\n@one\empty\else ^{\box\z@}\fi
\ifx\n@two\empty\else _{\box\@ne}\fi}}
\def\tw@@^#1{\@ifnextchar_ {\two@^{#1}}{\two@^{#1}_{}}}
\def\two{\@ifnextchar^ {\tw@@}{\tw@@^{}}}

\def\tofr@^#1_#2{\def\n@one{#1}\def\n@two{#2}\mathrel{\setw@dth{#1}{#2}
\mathop{\vcenter{\hbox to \w@dth{\rightarrowfill}\kern-1.7ex
                 \hbox to \w@dth{\leftarrowfill}}%
       }\limits
\ifx\n@one\empty\else ^{\box\z@}\fi
\ifx\n@two\empty\else _{\box\@ne}\fi}}
\def\t@fr@^#1{\@ifnextchar_ {\tofr@^{#1}}{\tofr@^{#1}_{}}}
\def\tofro{\@ifnextchar^ {\t@fr@}{\t@fr@^{}}}

\def\mon{\mathop{\m@th\hbox to
      14.6\P@{\lasyb\char'51\hskip-2.1\P@$\arrext$\hss
$\mathord\rightarrow$}}\limits} 
\def\leftmono{\mathrel{\m@th\hbox to
14.6\P@{$\mathord\leftarrow$\hss$\arrext$\hskip-2.1\P@\lasyb\char'50%
}}\limits} 
\mathchardef\arrext="0200       

\def\settypes(#1,#2,#3){\arrowtypea#1 \arrowtypeb#2 \arrowtypec#3}
\def\settoheight#1#2{\setbox\@tempboxa\hbox{#2}#1\ht\@tempboxa\relax}%
\def\settodepth#1#2{\setbox\@tempboxa\hbox{#2}#1\dp\@tempboxa\relax}%
\def\settokens[#1`#2`#3`#4]{%
     \def\tokena{#1}\def\tokenb{#2}\def\tokenc{#3}\def\tokend{#4}}
\def\setsqparms[#1`#2`#3`#4;#5`#6]{%
\arrowtypea #1
\arrowtypeb #2
\arrowtypec #3
\arrowtyped #4
\width #5
\height #6
}
\def\setpos(#1,#2){\xpos=#1 \ypos#2}

\def\settriparms[#1`#2`#3;#4]{\settripairparms[#1`#2`#3`1`1;#4]}%

\def\settripairparms[#1`#2`#3`#4`#5;#6]{%
\arrowtypea #1
\arrowtypeb #2
\arrowtypec #3
\arrowtyped #4
\arrowtypee #5
\width #6
\height #6
}

\def\resetparms{\settripairparms[1`1`1`1`1;500]\width 500}

\resetparms

\def\mvector(#1,#2)#3{
\put(0,0){\vector(#1,#2){#3}}%
\put(0,0){\vector(#1,#2){26}}%
}
\def\evector(#1,#2)#3{{
\arrowlength #3
\put(0,0){\vector(#1,#2){\arrowlength}}%
\advance \arrowlength by-30
\put(0,0){\vector(#1,#2){\arrowlength}}%
}}

\def\horsize#1#2{%
\settowidth{\tempdimen}{$#2$}%
#1=\tempdimen
\divide #1 by\unitlength
}

\def\vertsize#1#2{%
\settoheight{\tempdimen}{$#2$}%
#1=\tempdimen
\settodepth{\tempdimen}{$#2$}%
\advance #1 by\tempdimen
\divide #1 by\unitlength
}

\def\putvector(#1,#2)(#3,#4)#5#6{{%
\ifnum3<\arrowtype
\putdashvector(#1,#2)(#3,#4)#5\arrowtype
\else
\ifnum\arrowtype<-3
\putdashvector(#1,#2)(#3,#4)#5\arrowtype
\else
\xpos=#1
\ypos=#2
\run=#3
\rise=#4
\arrowlength=#5
\ifnum \arrowtype<0
    \ifnum \run=0
        \advance \ypos by-\arrowlength
    \else
        \tempcounta \arrowlength
        \multiply \tempcounta by\rise
        \divide \tempcounta by\run
        \ifnum\run>0
            \advance \xpos by\arrowlength
            \advance \ypos by\tempcounta
        \else
            \advance \xpos by-\arrowlength
            \advance \ypos by-\tempcounta
        \fi
    \fi
    \multiply \arrowtype by-1
    \multiply \rise by-1
    \multiply \run by-1
\fi
\ifcase \arrowtype
\or \put(\xpos,\ypos){\vector(\run,\rise){\arrowlength}}%
\or \put(\xpos,\ypos){\mvector(\run,\rise)\arrowlength}%
\or \put(\xpos,\ypos){\evector(\run,\rise){\arrowlength}}%
\fi\fi\fi
}}

\def\putsplitvector(#1,#2)#3#4{
\xpos #1
\ypos #2
\arrowtype #4
\halflength #3
\arrowlength #3
\gap 140
\advance \halflength by-\gap
\divide \halflength by2
\ifnum\arrowtype>0
   \ifcase \arrowtype
   \or \put(\xpos,\ypos){\line(0,-1){\halflength}}%
       \advance\ypos by-\halflength
       \advance\ypos by-\gap
       \put(\xpos,\ypos){\vector(0,-1){\halflength}}%
   \or \put(\xpos,\ypos){\line(0,-1)\halflength}%
       \put(\xpos,\ypos){\vector(0,-1)3}%
       \advance\ypos by-\halflength
       \advance\ypos by-\gap
       \put(\xpos,\ypos){\vector(0,-1){\halflength}}%
   \or \put(\xpos,\ypos){\line(0,-1)\halflength}%
       \advance\ypos by-\halflength
       \advance\ypos by-\gap
       \put(\xpos,\ypos){\evector(0,-1){\halflength}}%
   \fi
\else \arrowtype=-\arrowtype
   \ifcase\arrowtype
   \or \advance \ypos by-\arrowlength
       \put(\xpos,\ypos){\line(0,1){\halflength}}%
       \advance\ypos by\halflength
       \advance\ypos by\gap
       \put(\xpos,\ypos){\vector(0,1){\halflength}}%
   \or \advance \ypos by-\arrowlength
       \put(\xpos,\ypos){\line(0,1)\halflength}%
       \put(\xpos,\ypos){\vector(0,1)3}%
       \advance\ypos by\halflength
       \advance\ypos by\gap
       \put(\xpos,\ypos){\vector(0,1){\halflength}}%
   \or \advance \ypos by-\arrowlength
       \put(\xpos,\ypos){\line(0,1)\halflength}%
       \advance\ypos by\halflength
       \advance\ypos by\gap
       \put(\xpos,\ypos){\evector(0,1){\halflength}}%
   \fi
\fi
}

\def\putmorphism(#1)(#2,#3)[#4`#5`#6]#7#8#9{{%
\run #2
\rise #3
\ifnum\rise=0
  \puthmorphism(#1)[#4`#5`#6]{#7}{#8}#9%
\else\ifnum\run=0
  \putvmorphism(#1)[#4`#5`#6]{#7}{#8}#9%
\else
\setpos(#1)%
\arrowlength #7
\arrowtype #8
\ifnum\run=0
\else\ifnum\rise=0
\else
\ifnum\run>0
    \coefa=1
\else
   \coefa=-1
\fi
\ifnum\arrowtype>0
   \coefb=0
   \coefc=-1
\else
   \coefb=\coefa
   \coefc=1
   \arrowtype=-\arrowtype
\fi
\width=2
\multiply \width by\run
\divide \width by\rise
\ifnum \width<0  \width=-\width\fi
\advance\width by60
\if l#9 \width=-\width\fi
\putbox(\xpos,\ypos){#4}
{\multiply \coefa by\arrowlength
\advance\xpos by\coefa
\multiply \coefa by\rise
\divide \coefa by\run
\advance \ypos by\coefa
\putbox(\xpos,\ypos){#5} }%
{\multiply \coefa by\arrowlength
\divide \coefa by2
\advance \xpos by\coefa
\advance \xpos by\width
\multiply \coefa by\rise
\divide \coefa by\run
\advance \ypos by\coefa
\if l#9%
   \putrbox(\xpos,\ypos){#6}%
\else\if r#9%
   \putlbox(\xpos,\ypos){#6}%
\fi\fi }%
{\multiply \rise by-\coefc
\multiply \run by-\coefc
\multiply \coefb by\arrowlength
\advance \xpos by\coefb
\multiply \coefb by\rise
\divide \coefb by\run
\advance \ypos by\coefb
\multiply \coefc by70
\advance \ypos by\coefc
\multiply \coefc by\run
\divide \coefc by\rise
\advance \xpos by\coefc
\multiply \coefa by140
\multiply \coefa by\run
\divide \coefa by\rise
\advance \arrowlength by\coefa
\ifcase\arrowtype
\or \put(\xpos,\ypos){\vector(\run,\rise){\arrowlength}}%
\or \put(\xpos,\ypos){\mvector(\run,\rise){\arrowlength}}%
\or \put(\xpos,\ypos){\evector(\run,\rise){\arrowlength}}%
\fi}\fi\fi\fi\fi}}

\newcount\numbdashes \newcount\lengthdash \newcount\increment

\def\howmanydashes{
\numbdashes=\arrowlength \lengthdash=40
\divide\numbdashes by \lengthdash
\lengthdash=\arrowlength
\divide\lengthdash by \numbdashes
\increment=\lengthdash
\multiply\lengthdash by 3
\divide\lengthdash by 5
}

\def\putdashvector(#1)(#2,#3)#4#5{%
\ifnum#3=0 \putdashhvector(#1){#4}#5
\else
\ifnum#2=0
\putdashvvector(#1){#4}#5\fi\fi}

\def\putdashhvector(#1,#2)#3#4{{%
\arrowlength=#3 \howmanydashes
\multiput(#1,#2)(\increment,0){\numbdashes}%
{\vrule height .4pt width \lengthdash\unitlength}
\arrowtype=#4 \xpos=#1
\ifnum\arrowtype<0 \advance\arrowtype by 7 \fi
\ifcase\arrowtype
\or \advance\xpos by 10
    \put(\xpos,#2){\vector(-1,0){\lengthdash}}
    \advance\xpos by 40
    \put(\xpos,#2){\vector(-1,0){\lengthdash}}
\or \advance \xpos by 10
    \put(\xpos,#2){\vector(-1,0){\lengthdash}}
    \advance\xpos by  \arrowlength
    \advance\xpos by  -50
    \put(\xpos,#2){\vector(-1,0){\lengthdash}}
\or \advance\xpos by 10
    \put(\xpos,#2){\vector(-1,0){\lengthdash}}
\or \advance\xpos by \arrowlength
    \advance\xpos by -\lengthdash
    \put(\xpos,#2){\vector(1,0){\lengthdash}}
\or {\advance\xpos by 10
    \put(\xpos,#2){\vector(1,0){\lengthdash}}}
    \advance\xpos by \arrowlength
    \advance\xpos by -\lengthdash
    \put(\xpos,#2){\vector(1,0){\lengthdash}}
\or \advance\xpos by \arrowlength
    \advance\xpos by -\lengthdash
    \put(\xpos,#2){\vector(1,0){\lengthdash}}
    \advance\xpos by -40
    \put(\xpos,#2){\vector(1,0){\lengthdash}}
   \fi
}}

\def\putdashvvector(#1,#2)#3#4{{%
\arrowlength=#3 \howmanydashes
\ypos=#2 \advance\ypos by -\arrowlength
\multiput(#1,#2)(0,\increment){\numbdashes}%
    {\vrule width .4pt height \lengthdash\unitlength}
\arrowtype=#4 \ypos=#2
\ifnum\arrowtype<0 \advance\arrowtype by 7 \fi
\ifcase\arrowtype
\or \advance\ypos by \arrowlength \advance\ypos by -40
    \put(#1,\ypos){\vector(0,1){\lengthdash}}
    \advance\ypos by -40
    \put(#1,\ypos){\vector(0,1){\lengthdash}}
\or \advance\ypos by 10
    \put(#1,\ypos){\vector(0,1){\lengthdash}}
    \advance\ypos by \arrowlength \advance\ypos by -40
    \put(#1,\ypos){\vector(0,1){\lengthdash}}
\or \advance\ypos by \arrowlength \advance\ypos by -40
    \put(#1,\ypos){\vector(0,1){\lengthdash}}
\or \advance\ypos by 10
    \put(#1,\ypos){\vector(0,-1){\lengthdash}}
\or \advance\ypos by 10
    \put(#1,\ypos){\vector(0,-1){\lengthdash}}
    \advance\ypos by \arrowlength \advance\ypos by -40
    \put(#1,\ypos){\vector(0,-1){\lengthdash}}
\or \advance\ypos by 10
    \put(#1,\ypos){\vector(0,-1){\lengthdash}}
    \advance\ypos by 40
    \put(#1,\ypos){\vector(0,-1){\lengthdash}}
\fi
}}

\def\puthmorphism(#1,#2)[#3`#4`#5]#6#7#8{{%
\xpos #1
\ypos #2
\width #6
\arrowlength #6
\arrowtype=#7
\putbox(\xpos,\ypos){#3\vphantom{#4}}%
{\advance \xpos by\arrowlength
\putbox(\xpos,\ypos){\vphantom{#3}#4}}%
\horsize{\tempcounta}{#3}%
\horsize{\tempcountb}{#4}%
\divide \tempcounta by2
\divide \tempcountb by2
\advance \tempcounta by30
\advance \tempcountb by30
\advance \xpos by\tempcounta
\advance \arrowlength by-\tempcounta
\advance \arrowlength by-\tempcountb
\putvector(\xpos,\ypos)(1,0)\arrowlength\arrowtype
\divide \arrowlength by2
\advance \xpos by\arrowlength
\vertsize{\tempcounta}{#5}%
\divide\tempcounta by2
\advance \tempcounta by20
\if a#8 %
   \advance \ypos by\tempcounta
   \putbox(\xpos,\ypos){#5}%
\else
   \advance \ypos by-\tempcounta
   \putbox(\xpos,\ypos){#5}%
\fi}}

\def\putvmorphism(#1,#2)[#3`#4`#5]#6#7#8{{%
\xpos #1
\ypos #2
\arrowlength #6
\arrowtype #7
\settowidth{\xlen}{$#5$}%
\putbox(\xpos,\ypos){#3}%
{\advance \ypos by-\arrowlength
\putbox(\xpos,\ypos){#4}}%
{\advance\arrowlength by-140
\advance \ypos by-70
\ifdim\xlen>0pt
   \if m#8%
      \putsplitvector(\xpos,\ypos)\arrowlength\arrowtype
   \else
   \putvector(\xpos,\ypos)(0,-1)\arrowlength\arrowtype
   \fi
\else
   \putvector(\xpos,\ypos)(0,-1)\arrowlength\arrowtype
\fi}%
\ifdim\xlen>0pt
   \divide \arrowlength by2
   \advance\ypos by-\arrowlength
   \if l#8%
      \advance \xpos by-40
      \putrbox(\xpos,\ypos){#5}%
   \else\if r#8%
      \advance \xpos by40
      \putlbox(\xpos,\ypos){#5}%
   \else
      \putbox(\xpos,\ypos){#5}%
   \fi\fi
\fi
}}

\def\putsquarep<#1>(#2)[#3;#4`#5`#6`#7]{{%
\setsqparms[#1]%
\setpos(#2)%
\settokens[#3]%
\puthmorphism(\xpos,\ypos)[\tokenc`\tokend`{#7}]{\width}{\arrowtyped}b%
\advance\ypos by \height
\puthmorphism(\xpos,\ypos)[\tokena`\tokenb`{#4}]{\width}{\arrowtypea}a%
\putvmorphism(\xpos,\ypos)[``{#5}]{\height}{\arrowtypeb}l%
\advance\xpos by \width
\putvmorphism(\xpos,\ypos)[``{#6}]{\height}{\arrowtypec}r%
}}

\def\putsquare{\@ifnextchar <{\putsquarep}{\putsquarep%
   <\arrowtypea`\arrowtypeb`\arrowtypec`\arrowtyped;\width`\height>}}
\def\square{\@ifnextchar< {\squarep}{\squarep
   <\arrowtypea`\arrowtypeb`\arrowtypec`\arrowtyped;\width`\height>}}
\def\squarep<#1>[#2`#3`#4`#5;#6`#7`#8`#9]{{
\setsqparms[#1]
\diagram
\putsquarep<\arrowtypea`\arrowtypeb`\arrowtypec`
\arrowtyped;\width`\height>
(0,0)[#2`#3`#4`{#5};#6`#7`#8`{#9}]
\enddiagram
}}                                                 
\def\putptrianglep<#1>(#2,#3)[#4`#5`#6;#7`#8`#9]{{%
\settriparms[#1]%
\xpos=#2 \ypos=#3
\advance\ypos by \height
\puthmorphism(\xpos,\ypos)[#4`#5`{#7}]{\height}{\arrowtypea}a%
\putvmorphism(\xpos,\ypos)[`#6`{#8}]{\height}{\arrowtypeb}l%
\advance\xpos by\height
\putmorphism(\xpos,\ypos)(-1,-1)[``{#9}]{\height}{\arrowtypec}r%
}}

\def\putptriangle{\@ifnextchar <{\putptrianglep}{\putptrianglep
   <\arrowtypea`\arrowtypeb`\arrowtypec;\height>}}
\def\ptriangle{\@ifnextchar <{\ptrianglep}{\ptrianglep
   <\arrowtypea`\arrowtypeb`\arrowtypec;\height>}}
\def\ptrianglep<#1>[#2`#3`#4;#5`#6`#7]{{
\settriparms[#1]
\diagram
\putptrianglep<\arrowtypea`\arrowtypeb`
\arrowtypec;\height>
(0,0)[#2`#3`#4;#5`#6`{#7}]
\enddiagram
}}                                            

\def\putqtrianglep<#1>(#2,#3)[#4`#5`#6;#7`#8`#9]{{%
\settriparms[#1]%
\xpos=#2 \ypos=#3
\advance\ypos by\height
\puthmorphism(\xpos,\ypos)[#4`#5`{#7}]{\height}{\arrowtypea}a%
\putmorphism(\xpos,\ypos)(1,-1)[``{#8}]{\height}{\arrowtypeb}l%
\advance\xpos by\height
\putvmorphism(\xpos,\ypos)[`#6`{#9}]{\height}{\arrowtypec}r%
}}

\def\putqtriangle{\@ifnextchar <{\putqtrianglep}{\putqtrianglep
   <\arrowtypea`\arrowtypeb`\arrowtypec;\height>}}
\def\qtriangle{\@ifnextchar <{\qtrianglep}{\qtrianglep
   <\arrowtypea`\arrowtypeb`\arrowtypec;\height>}}
\def\qtrianglep<#1>[#2`#3`#4;#5`#6`#7]{{
\settriparms[#1]
\width=\height                                
\diagram
\putqtrianglep<\arrowtypea`\arrowtypeb`
\arrowtypec;\height>
(0,0)[#2`#3`#4;#5`#6`{#7}]
\enddiagram
}}

\def\putdtrianglep<#1>(#2,#3)[#4`#5`#6;#7`#8`#9]{{%
\settriparms[#1]%
\xpos=#2 \ypos=#3
\puthmorphism(\xpos,\ypos)[#5`#6`{#9}]{\height}{\arrowtypec}b%
\advance\xpos by \height \advance\ypos by\height
\putmorphism(\xpos,\ypos)(-1,-1)[``{#7}]{\height}{\arrowtypea}l%
\putvmorphism(\xpos,\ypos)[#4``{#8}]{\height}{\arrowtypeb}r%
}}

\def\putdtriangle{\@ifnextchar <{\putdtrianglep}{\putdtrianglep
   <\arrowtypea`\arrowtypeb`\arrowtypec;\height>}}
\def\dtriangle{\@ifnextchar <{\dtrianglep}{\dtrianglep
   <\arrowtypea`\arrowtypeb`\arrowtypec;\height>}}
\def\dtrianglep<#1>[#2`#3`#4;#5`#6`#7]{{
\settriparms[#1]
\width=\height                                
\diagram
\putdtrianglep<\arrowtypea`\arrowtypeb`
\arrowtypec;\height>
(0,0)[#2`#3`#4;#5`#6`{#7}]
\enddiagram
}}

\def\putbtrianglep<#1>(#2,#3)[#4`#5`#6;#7`#8`#9]{{%
\settriparms[#1]%
\xpos=#2 \ypos=#3
\puthmorphism(\xpos,\ypos)[#5`#6`{#9}]{\height}{\arrowtypec}b%
\advance\ypos by\height
\putmorphism(\xpos,\ypos)(1,-1)[``{#8}]{\height}{\arrowtypeb}r%
\putvmorphism(\xpos,\ypos)[#4``{#7}]{\height}{\arrowtypea}l%
}}

\def\putbtriangle{\@ifnextchar <{\putbtrianglep}{\putbtrianglep
   <\arrowtypea`\arrowtypeb`\arrowtypec;\height>}}
\def\btriangle{\@ifnextchar <{\btrianglep}{\btrianglep
   <\arrowtypea`\arrowtypeb`\arrowtypec;\height>}}
\def\btrianglep<#1>[#2`#3`#4;#5`#6`#7]{{
\settriparms[#1]
\width=\height                               
\diagram
\putbtrianglep<\arrowtypea`\arrowtypeb`
\arrowtypec;\height>
(0,0)[#2`#3`#4;#5`#6`{#7}]
\enddiagram
}}

\def\putAtrianglep<#1>(#2,#3)[#4`#5`#6;#7`#8`#9]{{%
\settriparms[#1]%
\xpos=#2 \ypos=#3
{\multiply \height by2
\puthmorphism(\xpos,\ypos)[#5`#6`{#9}]{\height}{\arrowtypec}b}%
\advance\xpos by\height \advance\ypos by\height
\putmorphism(\xpos,\ypos)(-1,-1)[#4``{#7}]{\height}{\arrowtypea}l%
\putmorphism(\xpos,\ypos)(1,-1)[``{#8}]{\height}{\arrowtypeb}r%
}}

\def\putAtriangle{\@ifnextchar <{\putAtrianglep}{\putAtrianglep
   <\arrowtypea`\arrowtypeb`\arrowtypec;\height>}}
\def\Atriangle{\@ifnextchar <{\Atrianglep}{\Atrianglep
   <\arrowtypea`\arrowtypeb`\arrowtypec;\height>}}
\def\Atrianglep<#1>[#2`#3`#4;#5`#6`#7]{{
\settriparms[#1]
\width=\height                                     
\diagram
\putAtrianglep<\arrowtypea`\arrowtypeb`
\arrowtypec;\height>
(0,0)[#2`#3`#4;#5`#6`{#7}]
\enddiagram
}}

\def\putAtrianglepairp<#1>(#2)[#3;#4`#5`#6`#7`#8]{{%
\settripairparms[#1]%
\setpos(#2)%
\settokens[#3]%
\puthmorphism(\xpos,\ypos)[\tokenb`\tokenc`{#7}]{\height}{\arrowtyped}b%
\advance\xpos by\height
\puthmorphism(\xpos,\ypos)[\phantom{\tokenc}`\tokend`{#8}]%
{\height}{\arrowtypee}b%
\advance\ypos by\height
\putmorphism(\xpos,\ypos)(-1,-1)[\tokena``{#4}]{\height}{\arrowtypea}l%
\putvmorphism(\xpos,\ypos)[``{#5}]{\height}{\arrowtypeb}m%
\putmorphism(\xpos,\ypos)(1,-1)[``{#6}]{\height}{\arrowtypec}r%
}}

\def\putAtrianglepair{\@ifnextchar <{\putAtrianglepairp}{\putAtrianglepairp%
   <\arrowtypea`\arrowtypeb`\arrowtypec`\arrowtyped`\arrowtypee;\height>}}
\def\Atrianglepair{\@ifnextchar <{\Atrianglepairp}{\Atrianglepairp%
   <\arrowtypea`\arrowtypeb`\arrowtypec`\arrowtyped`\arrowtypee;\height>}}

\def\Atrianglepairp<#1>[#2;#3`#4`#5`#6`#7]{{
\settripairparms[#1]
\settokens[#2]
\width=\height                                
\diagram
\putAtrianglepairp                            
<\arrowtypea`\arrowtypeb`\arrowtypec`
\arrowtyped`\arrowtypee;\height>
(0,0)[{#2};#3`#4`#5`#6`{#7}]
\enddiagram
}}

\def\putVtrianglep<#1>(#2,#3)[#4`#5`#6;#7`#8`#9]{{%
\settriparms[#1]%
\xpos=#2 \ypos=#3
\advance\ypos by\height
{\multiply\height by2
\puthmorphism(\xpos,\ypos)[#4`#5`{#7}]{\height}{\arrowtypea}a}%
\putmorphism(\xpos,\ypos)(1,-1)[`#6`{#8}]{\height}{\arrowtypeb}l%
\advance\xpos by\height
\advance\xpos by\height
\putmorphism(\xpos,\ypos)(-1,-1)[``{#9}]{\height}{\arrowtypec}r%
}}

\def\putVtriangle{\@ifnextchar <{\putVtrianglep}{\putVtrianglep
   <\arrowtypea`\arrowtypeb`\arrowtypec;\height>}}
\def\Vtriangle{\@ifnextchar <{\Vtrianglep}{\Vtrianglep
   <\arrowtypea`\arrowtypeb`\arrowtypec;\height>}}
\def\Vtrianglep<#1>[#2`#3`#4;#5`#6`#7]{{
\settriparms[#1]
\width=\height                                 
\diagram
\putVtrianglep<\arrowtypea`\arrowtypeb`
\arrowtypec;\height>
(0,0)[#2`#3`#4;#5`#6`{#7}]
\enddiagram
}}

\def\putVtrianglepairp<#1>(#2)[#3;#4`#5`#6`#7`#8]{{
\settripairparms[#1]%
\setpos(#2)%
\settokens[#3]%
\advance\ypos by\height
\putmorphism(\xpos,\ypos)(1,-1)[`\tokend`{#6}]{\height}{\arrowtypec}l%
\puthmorphism(\xpos,\ypos)[\tokena`\tokenb`{#4}]{\height}{\arrowtypea}a%
\advance\xpos by\height
\puthmorphism(\xpos,\ypos)[\phantom{\tokenb}`\tokenc`{#5}]%
{\height}{\arrowtypeb}a%
\putvmorphism(\xpos,\ypos)[``{#7}]{\height}{\arrowtyped}m%
\advance\xpos by\height
\putmorphism(\xpos,\ypos)(-1,-1)[``{#8}]{\height}{\arrowtypee}r%
}}

\def\putVtrianglepair{\@ifnextchar <{\putVtrianglepairp}{\putVtrianglepairp%
    <\arrowtypea`\arrowtypeb`\arrowtypec`\arrowtyped`\arrowtypee;\height>}}
\def\Vtrianglepair{\@ifnextchar <{\Vtrianglepairp}{\Vtrianglepairp%
    <\arrowtypea`\arrowtypeb`\arrowtypec`\arrowtyped`\arrowtypee;\height>}}
\def\Vtrianglepairp<#1>[#2;#3`#4`#5`#6`#7]{{
\settripairparms[#1]
\settokens[#2]
\diagram
\putVtrianglepairp                             
<\arrowtypea`\arrowtypeb`\arrowtypec`
\arrowtyped`\arrowtypee;\height>
(0,0)[{#2};#3`#4`#5`#6`{#7}]
\enddiagram
}}

\def\putCtrianglep<#1>(#2,#3)[#4`#5`#6;#7`#8`#9]{{%
\settriparms[#1]%
\xpos=#2 \ypos=#3
\advance\ypos by\height
\putmorphism(\xpos,\ypos)(1,-1)[``{#9}]{\height}{\arrowtypec}l%
\advance\xpos by\height
\advance\ypos by\height
\putmorphism(\xpos,\ypos)(-1,-1)[#4`#5`{#7}]{\height}{\arrowtypea}l%
{\multiply\height by 2
\putvmorphism(\xpos,\ypos)[`#6`{#8}]{\height}{\arrowtypeb}r}%
}}

\def\putCtriangle{\@ifnextchar <{\putCtrianglep}{\putCtrianglep
    <\arrowtypea`\arrowtypeb`\arrowtypec;\height>}}
\def\Ctriangle{\@ifnextchar <{\Ctrianglep}{\Ctrianglep
    <\arrowtypea`\arrowtypeb`\arrowtypec;\height>}}
\def\Ctrianglep<#1>[#2`#3`#4;#5`#6`#7]{{
\settriparms[#1]
\width=\height                               
\diagram
\putCtrianglep<\arrowtypea`\arrowtypeb`
\arrowtypec;\height>
(0,0)[#2`#3`#4;#5`#6`{#7}]
\enddiagram
}}                                           
\def\putDtrianglep<#1>(#2,#3)[#4`#5`#6;#7`#8`#9]{{%
\settriparms[#1]%
\xpos=#2 \ypos=#3
\advance\xpos by\height \advance\ypos by\height
\putmorphism(\xpos,\ypos)(-1,-1)[``{#9}]{\height}{\arrowtypec}r%
\advance\xpos by-\height \advance\ypos by\height
\putmorphism(\xpos,\ypos)(1,-1)[`#5`{#8}]{\height}{\arrowtypeb}r%
{\multiply\height by 2
\putvmorphism(\xpos,\ypos)[#4`#6`{#7}]{\height}{\arrowtypea}l}%
}}

\def\putDtriangle{\@ifnextchar <{\putDtrianglep}{\putDtrianglep
    <\arrowtypea`\arrowtypeb`\arrowtypec;\height>}}
\def\Dtriangle{\@ifnextchar <{\Dtrianglep}{\Dtrianglep
   <\arrowtypea`\arrowtypeb`\arrowtypec;\height>}}
\def\Dtrianglep<#1>[#2`#3`#4;#5`#6`#7]{{
\settriparms[#1]
\width=\height                              
\diagram
\putDtrianglep<\arrowtypea`\arrowtypeb`
\arrowtypec;\height>
(0,0)[#2`#3`#4;#5`#6`{#7}]
\enddiagram
}}                                          
\def\setrecparms[#1`#2]{\width=#1 \height=#2}%

\def\recursep<#1`#2>[#3;#4`#5`#6`#7`#8]{{%
\width=#1 \height=#2
\settokens[#3]
\settowidth{\tempdimen}{$\tokena$}
\ifdim\tempdimen=0pt
  \savebox{\tempboxa}{\hbox{$\tokenb$}}%
  \savebox{\tempboxb}{\hbox{$\tokend$}}%
  \savebox{\tempboxc}{\hbox{$#6$}}%
\else
  \savebox{\tempboxa}{\hbox{$\hbox{$\tokena$}\times\hbox{$\tokenb$}$}}%
  \savebox{\tempboxb}{\hbox{$\hbox{$\tokena$}\times\hbox{$\tokend$}$}}%
  \savebox{\tempboxc}{\hbox{$\hbox{$\tokena$}\times\hbox{$#6$}$}}%
\fi
\ypos=\height
\divide\ypos by 2
\xpos=\ypos
\advance\xpos by \width
\bfig
\putCtrianglep<-1`1`1;\ypos>(0,0)[`\tokenc`;#5`#6`{#7}]%
\puthmorphism(\ypos,0)[\tokend`\usebox{\tempboxb}`{#8}]{\width}{-1}b%
\puthmorphism(\ypos,\height)[\tokenb`\usebox{\tempboxa}`{#4}]{\width}{-1}a%
\advance\ypos by \width
\putvmorphism(\ypos,\height)[``\usebox{\tempboxc}]{\height}1r%
\efig
}}

\def\recurse{\@ifnextchar <{\recursep}{\recursep<\width`\height>}}

\def\puttwohmorphisms(#1,#2)[#3`#4;#5`#6]#7#8#9{{%
\puthmorphism(#1,#2)[#3`#4`]{#7}0a
\ypos=#2
\advance\ypos by 20
\puthmorphism(#1,\ypos)[\phantom{#3}`\phantom{#4}`#5]{#7}{#8}a
\advance\ypos by -40
\puthmorphism(#1,\ypos)[\phantom{#3}`\phantom{#4}`#6]{#7}{#9}b
}}

\def\puttwovmorphisms(#1,#2)[#3`#4;#5`#6]#7#8#9{{%
\putvmorphism(#1,#2)[#3`#4`]{#7}0a
\xpos=#1
\advance\xpos by -20
\putvmorphism(\xpos,#2)[\phantom{#3}`\phantom{#4}`#5]{#7}{#8}l
\advance\xpos by 40
\putvmorphism(\xpos,#2)[\phantom{#3}`\phantom{#4}`#6]{#7}{#9}r
}}

\def\puthcoequalizer(#1)[#2`#3`#4;#5`#6`#7]#8#9{{%
\setpos(#1)%
\puttwohmorphisms(\xpos,\ypos)[#2`#3;#5`#6]{#8}11%
\advance\xpos by #8
\puthmorphism(\xpos,\ypos)[\phantom{#3}`#4`#7]{#8}1{#9}
}}

\def\putvcoequalizer(#1)[#2`#3`#4;#5`#6`#7]#8#9{{%
\setpos(#1)%
\puttwovmorphisms(\xpos,\ypos)[#2`#3;#5`#6]{#8}11%
\advance\ypos by -#8
\putvmorphism(\xpos,\ypos)[\phantom{#3}`#4`#7]{#8}1{#9}
}}

\def\putthreehmorphisms(#1)[#2`#3;#4`#5`#6]#7(#8)#9{{%
\setpos(#1) \settypes(#8)
\if a#9 %
     \vertsize{\tempcounta}{#5}%
     \vertsize{\tempcountb}{#6}%
     \ifnum \tempcounta<\tempcountb \tempcounta=\tempcountb \fi
\else
     \vertsize{\tempcounta}{#4}%
     \vertsize{\tempcountb}{#5}%
     \ifnum \tempcounta<\tempcountb \tempcounta=\tempcountb \fi
\fi
\advance \tempcounta by 60
\puthmorphism(\xpos,\ypos)[#2`#3`#5]{#7}{\arrowtypeb}{#9}
\advance\ypos by \tempcounta
\puthmorphism(\xpos,\ypos)[\phantom{#2}`\phantom{#3}`#4]{#7}{\arrowtypea}{#9}
\advance\ypos by -\tempcounta \advance\ypos by -\tempcounta
\puthmorphism(\xpos,\ypos)[\phantom{#2}`\phantom{#3}`#6]{#7}{\arrowtypec}{#9}
}}

\def\setarrowtoks[#1`#2`#3`#4`#5`#6]{%
\def\toka{#1}
\def\tokb{#2}
\def\tokc{#3}
\def\tokd{#4}
\def\toke{#5}
\def\tokf{#6}
}
\def\hex{\@ifnextchar <{\hexp}{\hexp<1000`400>}}
\def\hexp<#1`#2>[#3`#4`#5`#6`#7`#8;#9]{%
\setarrowtoks[#9]
\yext=#2 \advance \yext by #2
\xext=#1 \advance\xext by \yext
\bfig
\putCtriangle<-1`0`1;#2>(0,0)[`#5`;\tokb``\tokd]
\xext=#1 \yext=#2 \advance \yext by #2
\putsquare<1`0`0`1;\xext`\yext>(#2,0)[#3`#4`#7`#8;\toka```\tokf]
\advance \xext by #2
\putDtriangle<0`1`-1;#2>(\xext,0)[`#6`;`\tokc`\toke]
\efig
}

\def\sAA{{\rm A\kern-0.85em A}} 
\def\tAA{{\mathchoice
  {\sAA}
  {\sAA}
  {\rm A\kern-0.60em A}
  {\rm A\kern-0.50em A} }}
\def\sBB{{\rm I\kern-.17em{}B}}
\def\BB{{\mathchoice
  {\sBB}
  {\sBB}
  {\rm I\kern-.13em{}B}
  {\rm I\kern-.13em{}B} }}
\def\sCC{{\kern 0.27em\vrule height1.45ex width0.03em depth0em
          \kern-0.30em\rm C}}
\def\CC{{\mathchoice
  {\sCC}
  {\sCC}
  {\kern 0.225em \vrule height1.05ex width0.025em depth0em \kern-0.25em \rm C}
  {\kern 0.180em \vrule height0.78ex width0.02em depth0em \kern-0.2em \rm C}
        }}
\def\tCC{{\ooalign{C\crcr\kern0.27em\vrule height1.45ex width0.03em
depth0em\crcr}}}
\def\sDD{{\rm I\kern-.16em{}D}}
\def\DD{{\mathchoice
  {\sDD}
  {\sDD}
  {\rm I\kern-.13em{}D}
  {\rm I\kern-.13em{}D} }}
\def\sEE{{\rm I\kern-.17em{}E}}
\def\EE{{\mathchoice
  {\sEE}
  {\sEE}
  {\rm I\kern-.13em{}E}
  {\rm I\kern-.13em{}E} }}
\def\sFF{{\rm I\kern-.16em{}F}}
\def\FF{{\mathchoice
  {\sFF}
  {\sFF}
  {\rm I\kern-.13em{}F}
  {\rm I\kern-.13em{}F} }}
\def\sGG{{\kern 0.27em \vrule height1.45ex width0.03em depth0em
          \kern-0.30em \rm G}}
\def\GG{{\mathchoice
  {\sGG}
  {\sGG}
  {\kern 0.225em \vrule height1.05ex width0.025em depth0em \kern-0.25em \rm G}
  {\kern 0.180em \vrule height0.78ex width0.020em depth0em \kern-0.20em \rm G}
        }}
\def\sHH{{\rm I\kern-.16em{}H}}
\def\HH{{\mathchoice
  {\sHH}
  {\sHH}
  {\rm I\kern-.13em{}H}
  {\rm I\kern-.13em{}H} }}
\def\sII{{\rm I\kern-.16em{}I}}
\def\II{{\mathchoice
  {\sII}
  {\sII}
  {\rm I\kern-.12em{}I}
  {\rm I\kern-.10em{}I} }}
\def\sJJ{{\kern0.17em\vrule height1.5ex width 0.03em depth0em
          \kern-.20em\rm J}}
\def\JJ{{\mathchoice
  {\sJJ}
  {\sJJ}
  {\kern0.150em\vrule height1.05ex width 0.025em depth0em\kern-.175em\rm J}
  {\kern0.135em\vrule height0.78ex width 0.020em depth0em\kern-.155em\rm J} }}
\def\sKK{{\rm I\kern-.16em{}K}}
\def\KK{{\mathchoice
  {\sKK}
  {\sKK}
  {\rm I\kern-.12em{}K}
  {\rm I\kern-.10em{}K} }}
\def\sLL{{\rm I\kern-.16em{}L}}
\def\LL{{\mathchoice
  {\sLL}
  {\sLL}
  {\rm I\kern-.12em{}L}
  {\rm I\kern-.10em{}L} }}
\def\sMM{{\rm I\kern-.16em{}M}}
\def\MM{{\mathchoice
  {\sMM}
  {\sMM}
  {\rm I\kern-.12em{}M}
  {\rm I\kern-.10em{}M} }}
\def\sNN{{\rm I\kern-.16em{}N}}
\def\NN{{\mathchoice
  {\sNN}
  {\sNN}
  {\rm I\kern-.12em{}N}
  {\rm I\kern-.10em{}N} }}
\def\sOO{{\kern 0.27em \vrule height1.50ex width0.03em depth0em
					\kern-0.30em \rm O}}
\def\OO{{\mathchoice
  {\sOO}
  {\sOO}
  {\kern 0.225em \vrule height1.05ex width0.025em depth0em \kern-0.25em \rm O}
  {\kern 0.180em \vrule height0.78ex width0.020em depth0em \kern-0.20em \rm O}
        }}
\def\sPP{{\rm I\kern-.16em{}P}}
\def\PP{{\mathchoice
  {\sPP}
  {\sPP}
  {\rm I\kern-.12em{}P}
  {\rm I\kern-.10em{}P} }}
\def\sQQ{{\kern 0.27em \vrule height1.45ex width0.03em depth0em
          \kern-0.30em \rm Q}}
\def\QQ{{\mathchoice
	{\sQQ}
	{\sQQ}
  {\kern 0.225em \vrule height1.05ex width0.025em depth0em \kern-0.25em \rm Q}
  {\kern 0.180em \vrule height0.78ex width0.020em depth0em \kern-0.20em \rm Q}
        }}
\def\sRR{{\rm I\kern-0.16em{}R}}
\def\RR{{\mathchoice
  {\sRR}
  {\sRR}
  {\rm I\kern-0.12em{}R}
  {\rm I\kern-0.10em{}R} }}
\def\sSS{{\rm S\kern-.45em{}S}}
\def\sTT{{\rm T\kern-.60em{}T}}
\def\TT{{\mathchoice
  {\sTT}
  {\sTT}
  {\rm T\kern-.45em{}T}
  {\rm T\kern-.38em{}T} }}
\def\sUU{{\rm U\kern-.60em{}U}}
\def\UU{{\mathchoice
  {\sUU}
  {\sUU}
  {\rm U\kern-.46em{}U}
  {\rm U\kern-.38em{}U} }}
\def\sVV{{\rm V\kern-.62em{}V}}
\def\VV{{\mathchoice
  {\sVV}
  {\sVV}
  {\rm V\kern-.46em{}V}
  {\rm V\kern-.38em{}V} }}
\def\sWW{{\rm W\kern-.92em{}W}}
\def\WW{{\mathchoice
  {\sWW}
  {\sWW}
  {\rm W\kern-.80em{}W}
  {\rm W\kern-.67em{}W} }}
\def\sXX{{\rm X\kern-.58em{}X}}
\def\XX{{\mathchoice
  {\sXX}
  {\sXX}
  {\rm X\kern-.45em{}X}
  {\rm X\kern-.38em{}X} }}
\def\sYY{{\rm Y\kern-.58em{}Y}}
\def\YY{{\mathchoice
  {\sYY}
  {\sYY}
  {\rm Y\kern-.45em{}Y}
  {\rm Y\kern-.40em{}Y} }}
\def\sZZ{{\rm Z\kern-0.32em{}Z}}
\def\ZZ{{\mathchoice
  {\sZZ}
  {\sZZ}
  {\rm Z\kern-0.30em{}Z}
  {\rm Z\kern-0.25em{}Z} }}

\begin{document}
\global\parskip=4pt


\begin{titlepage}
\newlength{\titlehead}
\settowidth{\titlehead}{NIKHEF-H/91}
\begin{flushright}
\parbox{\titlehead}{
\begin{flushleft}
IC/94/37\\
January 1994\\
\end{flushleft}
}
\end{flushright}
\begin{center}
\vskip .25in
{\LARGE\bf On Diagonalization in Map($\bf M,G$)} \\
\vskip .25in
{\bf Matthias Blau}\footnote{e-mail: blau@ictp.trieste.it}
and {\bf George Thompson}\footnote{e-mail: thompson@ictp.trieste.it}\\
\vskip .10in
ICTP\\
P.O. Box 586\\
I-34014 Trieste\\
Italy
\end{center}
\vskip .15in
\begin{abstract}
Motivated by some questions in the path integral approach to (topological)
gauge theories, we are led to address the following question: given a smooth
map from a manifold $M$
to a compact group $\bG$, is it possible to smoothly `diagonalize' it,
i.e.~conjugate it into a map to a maximal torus $\bT$ of $\bG$?

We analyze the local and global obstructions and give a complete solution to
the problem for regular maps. We establish that these can always be smoothly
diagonalized locally and that the obstructions to doing this globally are
non-trivial Weyl group and torus bundles on $M$. We show how the patching
of local
diagonalizing maps gives rise to non-trivial $\bT$-bundles, explain the
relation to winding numbers of maps into $\bG/\bT$ and restrictions of the
structure group and examine the behaviour of gauge fields under this
diagonalization. We also discuss the complications that arise for non-regular
maps and in the presence of non-trivial $\bG$-bundles. In particular, we
establish a relation between the existence of regular sections of a non-trivial
adjoint bundle and restrictions of the structure group of a principal
$\bG$-bundle to $\bT$.

We use these results to justify a Weyl integral formula for functional
integrals which, as a novel feature not seen in the finite-dimensional case,
contains a summation over all those topological $\bT$-sectors which arise
as restrictions of a trivial principal $\bG$ bundle and which was used
previously to solve completely Yang-Mills theory and the $G/G$ model in two
dimensions.

\end{abstract}
\end{titlepage}


\begin{small}
\tableofcontents
\end{small}

\setcounter{footnote}{0}

\section{Introduction}

One of the most useful properties of a compact Lie group $\bG$ is that
its elements can be `diagonalized' or, more formally, conjugated
into a fixed maximal torus $\bT\ss\bG$. In this paper
we investigate to which extent this property continues to hold for spaces
of smooth maps from a manifold $M$ to a compact Lie group $\bG$. Thus,
given a smooth map $g:M \ra \bG$, the first thing
one would like to know is if it can be written as
\be
g(x) = h^{-1}(x)t(x)h(x) \, , \label{conj1}
\ee
where $t:M \rightarrow \bT$ and $h:M \rightarrow \bG$ are smooth globally
defined maps. It is easy to see (by examples) that this cannot be true in
general, not even for loop groups ($M=S^{1}$),
and we are thus led to ask instead the following questions:
\begin{enumerate}
\item Under which conditions can (\ref{conj1}) be achieved locally on $M$?
\item Under which conditions will $t(x)$ be smooth (while possibly relaxing the
      conditions on $h$)?
\item What are the obstructions to representing $g$ as in (\ref{conj1})
      globally?
\end{enumerate}

We will not be able to answer these questions in full generality. For those
maps, however, which take values in the dense set $\bG_{r}$ of regular
elements of $\bG$ we provide complete answers to 1-3. We establish that
conjugation into $\bT$ can always be achieved locally
and that non-trivial $\bT$-bundles on $M$ are the obstructions to
finding smooth functions $h$ which accomplish (\ref{conj1}) globally.
Furthermore we prove that if either $\bG$ or $M$ is simply connected
the diagonalized map $t$ will be smooth globally. These results confirm the
intuition that (in $SU(n)$ language) obstructions to diagonalization can
arise from the ambiguities in either the phase of $h$ or in the ordering of
the eigenvalues of $t$.

While these questions seem to be interesting in their own right, they also
arise naturally within the context of gauge fixing in non-Abelian gauge
theories. In \cite{tho}, 't Hooft has argued that a `diagonalizing gauge'
may not only be technically useful but also essential for unravelling the
physical content of these theories. For us the
motivation for looking at this issue arose originally in the context
of low-dimensional gauge theories. In particular, in \cite{btver,btlec} we
used a path integral version of the Weyl integral formula, which relates the
integral of a conjugation invariant function over $\bG$ to an integral over
$\bT$, to effectively abelianize non-Abelian gauge theories like 2d Yang-Mills
theory and the $G/G$ gauged Wess-Zumino-Witten model. The path integrals
for the partition function and correlation functions on
arbitrary two-dimensional closed surfaces $\Sigma$ could then be calculated
explicitly and straightforwardly. Formally this Abelianization was achieved
by using the local conjugation (gauge) invariance of the action to impose the
`gauge condition' $g(x) \in \bT$ (or its Lie algebra counterpart in the case
of Yang-Mills theory). The correct results emerged when the resulting Abelian
theory was summed over all topological sectors of $\bT$-bundles on $\Sigma$,
even though the original $\bG$-bundle was trivial. This method has been
reviewed and applied to some other models recently in \cite{ew}.

In the light of the above, the occurrence of the sum over isomorphism
classes of $\bT$-bundles can now be understood as a consequence of the
fact that the chosen gauge condition cannot necessarily be achieved globally
on $M=\Sigma$ by smooth gauge transformations. But while it is certainly
legitimate to use a change of variables in the path integral which is not a
gauge transformation, one needs to exercise more care when keeping track of
the consequences of such a change of variables. Thus to the above list of
questions we add (with hindsight)
\begin{enumerate}
\setcounter{enumi}{3}
\item  What happens to $\bG$ gauge fields $A$ under the possibly non-smooth
       gauge transformation $A\ra A^{h} = h^{-1}A h  + h^{-1} dh$ ? In
       particular, does this give rise to $\bT$ gauge fields on non-trivial
       $\bT$ bundles on $M$?
\item What is the correct version of the path integral analogue of the Weyl
      integral formula taking into account the global obstructions to
      achieving (\ref{conj1}) globally? In particular, does this explain
      the appearance of the sum over all isomorphism classes of $\bT$ bundles?
\end{enumerate}
It turns out that indeed connections on $\bT$-bundles appear in that way and
that the Weyl integral formula should include a sum over those topological
sectors which appear as obstructions to diagonalization.

When $M$ and $\bG$
are such that there are no non-trivial $\bG$ bundles on $M$, all isomorphism
classes of torus bundles appear as obstructions (because then all torus bundles
are restrictions of the trivial $\bG$ bundle). In particular, modulo one
subtlety which we will come back to below, this takes care of the
two- and three-dimensional models considered in \cite{btver,btlec} (as
regular maps are generic in those cases and the contributions from the
non-regular maps are suppressed by the zeros of the Faddeev-Popov
determinant).

The close relation between
restrictions of the structure group of a principal $\bG$ bundle
and the existence of regular sections of its adjoint bundle
(which hence have smooth diagonalizations in the simply-connected case)
is expressed most clearly by our Proposition 6 which states that such
a restriction exists if and only if the adjoint bundle has a regular
section. This is, in a sense, the fundamental result of this paper.
Our proof relies on the previously established results concerning the
diagonalizability of maps. If one had a different and more direct proof of
this theorem (which, after all, does not refer explicitly to the issue of
conjugating maps into a maximal torus), then
the other existence results obtained in this paper could be obtained more
or less directly as Corollaries.

The situation concerning non-regular maps is much murkier and we will not
be able to say much about them. But we illustrate the difficulties which
arise in that case (and in the presence of non-trivial $\bG$ bundles)
by examples and discuss why and to which extent our present treatment
fails in these cases. The subtlety mentioned above arises because the
Wess-Zumino term in the $G/G$ model requires the extension of a $G$ valued
map $g$ to a bounding three-manifold and there are situations where the
extension is necessarily non-regular even if $g$ is regular. This problem
as well as the related issue of localization in the $G/G$ model are under
investigation at the moment \cite{btloc}.

This paper is organized as follows: In section 2 we briefly recall the basic
facts we need from the theory of Lie groups: maximal tori, the Weyl group,
universal and Weyl group coverings of the set of regular elements. In section 3
we discuss three prototypical examples which illustrate the possible ways in
which (\ref{conj1}) can fail either locally or globally. The first of these,
a smooth map from $S^{1}$ to $SU(2)$, shows that not even $t(x)$ is
necessarily smooth in general. The second, a regular map from $S^{2}$ to
$SU(2)$, can be smoothly diagonalized locally but not globally. It
provides a preliminary identification of certain obstructions in terms of
winding numbers of maps from $M$ to $\bG/\bT$ and also shows quite clearly how
and why connections on non-trivial $\bT$-bundles emerge. Finally. the
third example (a map into $SO(3)$) illustrates how global smoothness of $t$
can fail even for regular $g$ when both $M$ and $\bG$ are not simply connected.

Sections 4-6  contain the main mathematical results of this paper. In
section 4 we prove that regular maps can be smoothly conjugated into the
torus over any
contractible open set in $M$ and we identify the obstructions to doing this
globally. The results of this section are summarized in Propositions 1 and 2.
Proposition 3 contains the corresponding statements for Lie algebra valued
maps.

In section 5 we investigate what happens
when we try to patch together the local diagonalizing functions $h$
and rederive the previously found global obstructions from that point
of view. Focussing on the case when $\bG$ is simply
connected, we explain how finding a solution to (\ref{conj1}) is related
to restricting the structure group of a (trivial) principal $\bG$
bundle $P_{G}$ to $\bT$.
We also look at what happens to gauge fields on $P_{G}$ and explain the
relation between the Chern classes of non-trivial torus bundles on $M$ and the
winding numbers of maps from $M$ to $\bG/\bT$. One of the reasons why winding
numbers enter is because, in contrast with the space of maps from a
two-manifold into $\bG$,
the space of regular maps is not connected (Proposition 4).
These results can be captured
concisely by making use of an integral representation for a generalized winding
number of such maps, depending also on a $\bG$-connection $A$, and are
contained in Proposition 5 and the subsequent Corollaries.

In section 6 we return to those cases not covered by the previous analysis
and explain the complications which arise. In particular, we establish the
above-mentioned relation between the existence of
regular sections of the adjoint bundle ${\rm Ad}P_{G}$ of a non-trivial
$\bG$ bundle $P_{G}$ and restrictions of the structure group to $\bT$ or,
equivalently, sections of the associated $\bG/\bT$ bundle (Proposition 6).
We then discuss some higher dimensional examples which serve to illustrate
possible obstruction to restrictions of the structure group.
We also address the issue of genericity of regular maps and make some
(non-conclusive) comments on the problem of conjugating non-regular maps
into the torus.

Finally, in section 7, we turn to an applications of the above results.
We use them to justify a
version of the Weyl integral formula for functional integrals over
spaces of maps into a simply connected group. As a novel
feature not present in the finite dimensional (or quantum mechanical path
integral) version this formula includes a sum over all those topological
sectors of $\bT$ bundles which arise as restrictions of a trivial principal
$\bG$ bundle, justifying the method used in \cite{btver,btlec} to solve
exactly some low-dimensional (topological) gauge theories.

Although the entire paper has been phrased in the context of group valued maps,
most of it carries over, {\em mutatis mutandis}, to the case of Lie algebra
valued maps. We will point out as we go along whenever a non-obvious
difference arises between the two cases.

After having completed our investigations we came across a 1984 paper by
Grove and Pedersen \cite{gp} in which the local obstructions we find in
section 4 are also identified, albeit using quite different techniques,
see \cite[Theorem 1.4]{gp}.
The global issues which are our main concern in the present paper, in
particular the relation between conjugation into the torus and restrictions
of the structure group and the behaviour of gauge fields, are
not addressed in \cite{gp}, the emphasis there being on
characterizing those spaces on which every continuous
function taking values in normal matrices can be continuously diagonalized.
These turn out to be so-called sub-Stonean spaces of dimension $\leq 2$
satisfying certain additional criteria, \cite[Theorem 5.6]{gp}.

A final remark on terminology: we will (as above)
occasionally find it convenient to use $SU(n)$ terminology even when
dealing with a general compact Lie group $\bG$. Thus we might say `diagonalize'
when we should properly be saying `conjugate into the maximal torus' and
we may loosely refer to the action of the Weyl group as `a permutation of the
eigenvalues'. We denote the space of maps from a manifold $M$
into a group $\bG$ by $\mg$. Unless specified otherwise, these
maps are taken to be smooth.

\section{Background from the Theory of Lie Groups}

Let $\bG$ be a compact connected Lie group. A maximal torus $\bT$ of $\bG$ is
a maximal compact connected Abelian subgroup of $\bG$. Its dimension is
called the rank $r$ of $\bG$. Any two maximal tori of $\bG$ are conjugate
to each other, i.e.~if $\bT$ and $\bT'$ are maximal tori of $\bG$,
there exists a $h\in\bG$ such that $\bT' = h^{-1}\bT h$ and we will
henceforth choose one maximal torus arbitrarily and fix it. Since any
element of $\bG$ lies in some maximal torus, it follows that any element of
$\bG$ can be conjugated into $\bT$,
\be
\forall g \in \bG \; \exists h \in \bG : h g h^{-1} \in \bT\;\;.\label{g1}
\ee
Such a $h$ is of course not unique. First of all, $h$ can be multiplied
on the left by any element of $\bT$, $h \ra th, t \in \bT$ as $\bT$ is
Abelian. The residual conjugation
action of $\bG$ on $\bT$ (conjugation by elements of $\bG$ which leave $\bT$
invariant) is that of a finite group, the Weyl group
$W$. From the above description it follows that the Weyl group can be
thought of as the quotient $W=N(\bT)/\bT$, where
\be
N(\bT)= \{g\in\bG: \g t g\in \bT \;\forall t \in \bT\}
\ee
denotes the normalizer of $\bT$ in $\bG$. Thus
the complete ambiguity in $h$ satisfying (\ref{g1}) for a given $g$
is $h\ra nh, n\in N(\bT)$
and if $hgh^{-1} = t \in \bT$ then $(nh)g(nh)^{-1} = ntn^{-1} \in \bT$ is
one of the finite number of images $w(t)$ of $t$ under the action of the
Weyl group $W$.

For $\bG = SU(n)$, one has $\bT\sim U(1)^{n-1}$,
which can be realized by diagonal matrices in the fundamental
representation of $SU(n)$. $W$ is the permutation group $S_{n}$ on
$n$ objects
acting on an element of $\bT$ by permutation of the diagonal entries.

While it is true that any two maximal
tori are conjugate to each other, it is not necessarily
true that the
centralizer $C(g)$ of an element $g \in \bG$ (i.e.~the
set of elements of $\bG$ commuting with $g$) is some maximal torus.
For example, for $g$ an element of the center $Z(\bG)$ of $\bG$ one
obviously has $C(g)=\bG$. However, the set of elements of $\bG$
for which $\dim C(g) = \dim \bT$ is open and dense in $\bG$ and is called
the set $\bG_{r}$ of regular elements of $\bG$,
\be
\bG_{r}= \{g\in\bG: \dim C(g) =\dim \bT\}\;\;.
\ee
We also denote by $\bT_{r}$ the set of regular elements of $\bT$,
$\bT_{r}=\bT\cap\bG_{r}$. The regular elements of $\bG$ and $\bT$
can alternatively be characterized by the fact that they lie in one
and only one maximal torus of $\bG$ and this will give us some useful
information on the diagonalizability of regular maps in section 5.
Not only is $\bG_{r}$ open and dense in $\bG$ but the non-regular elements
actually form a set of codimension three in $\bG_{r}$. Although this set
may not be a manifold, $\bG_{r}$ and $\bG$ have the same fundamental group,
\be
\pi_{1}(\bG_{r})=\pi_{1}(\bG)\;\;.\label{pi1}
\ee

Even for a regular element $g\in\bG_{r}$, the
centralizer $C(g)$ need not be a maximal torus and hence conjugate to
$\bT$ (we will see an example of that below) and it will be convenient
to single out two further distinguished dense subsets of $\bT$ and (via
conjugation) $\bG$. We denote by $\bT_{n}$ and $\bT_{w}$ the sets of
elements $t$ of $\bT$ satisfying
\bea
\bT_{n}&=&\{t\in \bT: C(t)=\bT\}\;\;,\nonumber\\
\bT_{w}&=&\{t\in \bT: w(t)\neq t \;\;\forall w\in W, w\neq 1\}\;\;.
\eea
While perhaps not immediately evident, it is nevertheless true that
these two conditions are equivalent, $\bT_{n}=\bT_{w}$, so that,
as a consequence of the obvious inclusion $\bT_{n}\ss\bT_{r}$, one also has
$\bT_{w}\ss\bT_{r}$. Furthermore, if $\bG$ is simply connected,
$\pi_{1}(\bG)=0$, one has
\be
\pi_{1}(\bG)=0\Rightarrow \bT_{r}=\bT_{n}=\bT_{w}\;\;.
\ee

It can be shown (see e.g.~\cite{btd,hel}) that the conjugation map
\bea
q: \bG/\bT \times \bT_{r} &\ra& \bG_{r} \nonumber\\
([h],t) &\mapsto& h^{-1} t h \label{w1}
\eea
is a $|W|$-fold covering onto $\bG_{r}$ and that
$\gt\times\bT_{r}$ is the total space of a principal $W$ bundle over
$\bG_{r}$.

If $\bG$ is simply connected, (\ref{pi1}) implies that $\gt\times\bT_{r}$
is the total space of a
trival $W$-bundle over $\bG_{r}$ as any covering of $\bG_{r}$ is then
necessarily trivial. We will see in section 4 that this
simplifies the issue of diagonalizability of regular maps in that case.
It follows from the above that, for $\bG$ simply connected,
the Weyl group acts freely on each connected
component $\bP_{r}$ of $\bT_{r}=\bT_{w}$ and simply transitively on the set
of components.
Thus we can identify $\bP_{r}$, the image of a Weyl alcove under
the exponential map, with a fundamental domain $D$ for the action
of $W$ on $\bT_{r}$ and the restriction of $q$ to $\bP_{r}$ provides
an isomorphism between $\gt\times\bP_{r}$ and $\bG_{r}$. In particular,
as
\be
\pi_{2}(\gt)=\pi_{1}(\bT)=\ZZ^{r}\;\;,
\ee
this tells us that the second homotopy group of $\bG_{r}$ is
\be
\pi_{1}(\bG)=0\Rightarrow \pi_{2}(\bG_{r})=\ZZ^{r}\label{p2gr}
\ee
(to be contrasted with $\pi_{2}(\bG)=0$).
As the higher homotopy groups $\pi_{k}(\bT)$, $k>1$, are trivial,
it follows from the homotopy exact sequence associated to the fibration
$\bG\ra\gt$ that $\pi_{k}(\gt)=\pi_{k}(\bG)$ for $k>2$.
Thus by the same argument as above we can conclude that
\be
\pi_{1}(\bG)=0 \Rightarrow \pi_{k}(\bG_{r}) = \pi_{k}(\bG)\;\;\;\;\forall\;
k>2 \label{p3gr}\;\;.
\ee

In general, if one restricts $q$ to $\gt\times\bP_{r}$,
it becomes a universal covering of $\bG_{r}$. Thus $P_{r}$ will in
general contain points related by Weyl transfromations as well as fixed
points of $W$ as manifested by the fact that $\bT_{w}$ is not
necessarily equal to $\bT_{r}$ unless $\pi_{1}(\bG)=0$.
Therefore, for general compact $\bG$
the above covering (\ref{w1}) is neither trivial nor connected.
Nevertheless, the fact that, away from the non-regular points, the above map
$q$ is a smooth fibration (with discrete fibers) will be of utmost importance
in our discussion in section 4.

As an illustration of the above,
let us consider the groups $SU(2)$ and $SO(3)$. The only
non-regular elements of $SU(2)$ are $\pm{\bf 1}$.
Thus $SU(2)_{r}$ is isomorphic to a cylinder $S^{2}\times I$, $I=(0,2\pi)$
and one sees that $\pi_{2}(SU(2)_{r})=\pi_{3}(SU(2)_{r})=\ZZ$, in accordance
with (\ref{p2gr}) and (\ref{p3gr}).
The space $\bT_{r}$ of regular elements of the torus consists of two
connected components, $\bT_{r}\sim (0,\pi)\cup(\pi,2\pi)$,
which explains the triviality of the fibration (\ref{w1}) in that case.

$SO(3)_{r}$, on the other hand, is obtained from $SO(3)\sim\RR\PP^{3}$ by
removing one point, corresponding to the identity element.
The non-trivial double-covering $SU(2)\ra SO(3)$ restricts to the non-trivial
double covering $SU(2)_{r}\ra SO(3)_{r}$ and coincides with the fibration
(\ref{w1}) as $\bT_{r}(SO(3))\sim I$. The Weyl group acts on $\bT_{r}(SO(3))$
by $\varphi\ra 2\pi -\varphi$. Thus the point $\pi\in\bT_{r}(SO(3))$,
corresponding to the element $t={\rm diag}(-1,-1,1)$ in the standard
embedding of $\bT(SO(3))=SO(2)$ into $SO(3)$, is, while regular, a fixed
point of the Weyl group, $t\not\in\bT_{w}$. This is reflected in the fact
that the centralizer of
this element is $O(2)$, $t\not\in\bT_{n}$, which is strictly larger than than
$SO(2)$ while preserving the regularity condition $\dim C(t) = \dim \bT$.

\section{Examples: Obstructions to Globally Conjugating to the Torus}

We will now take a look at three examples of maps which illustrate the
obstructions to achieving (\ref{conj1}) globally or smoothly. The first
one, which we will only deal with briefly, illustrates what can go wrong
with maps which pass through non-regular points of $\bG$. We shall from
then on (and until section 6) focus exclusively on regular maps and try to
come to terms with them. The second example, a simple map from
$S^{2}$ to $SU(2)$, allows us to detect an obstruction to globally and
smoothly diagonalizing it more or less by inspection. This obstruction
turns out to be a winding number associated with that map. Refining that
winding number to include a gauge field contribution one can moreover read off
directly that any attempt to force the map into the torus by a possibly
non-smooth (discontinuous) $h$ wil give rise to non-trivial torus gauge fields.
The third example, a map from the circle to $SO(3)$, highlights another
obstruction which can only arise when neither $\bG$ nor $M$ is simply
connected.

\subs{Example 1: A Map from $S^{1}$ to $SU(2)$}

Let $f$ be any smooth $\RR$-valued function on the real line
such that $f(x+2\pi) = -f(x)$. Then the map
$g\in {\rm Map}(S^{1},SU(2))$ (the loop group of $SU(2)$)
defined by
\be
g(x) = \mat{\cos f(x)}{-ie^{-ix/2}\sin f(x)}{-ie^{ix/2}\sin f(x)}{\cos f(x)}
\ee
is single-valued, $g(x+2\pi) = g(x)$, and smooth. As $f$ is necessarily zero
somewhere, $g$ passes through the (non-regular) identity element. $g$ can be
diagonalized by a map $h$, $hgh^{-1} =t$, but neither $h$ nor $t$ are single
valued on $S^{1}$. For instance, $h$ can be chosen to be
\be
h(x) = \frac{1}{\sqrt{2}}\mat{e^{ix/2}}{-1}{1}{e^{-ix/2}}\;\;,
\ee
and $t$ turns out to be
\bea
t(x) &=& \mat{e^{if(x)}}{0}{0}{e^{-if(x)}}\;\;,\\
t(x+2\pi) &=& t^{-1}(x)\neq t(x)\;\;.
\eea
What happens here is that, upon going around the circle, $t(x)$ comes back
to itself only up to the action of the Weyl group, reflecting the ambiguity
$h\ra nh$ mentioned in section 2. Had $g$ been regular to start off with,
this ambiguity could have been consistently eliminated by giving a
particular ordering prescription for the diagonal elements. Such a presciption,
however, becomes ambiguous when two of the diagonal elements coincide
(as at the identity element of the group).

\subs{Example 2: A map from $S^{2}$ to $SU(2)$}

A nice example (suggested to us by E.~Witten) giving us a first idea of the
possible obstructions in the case of regular maps and
the role of non-trivial torus bundles is afforded by the
following map from the sphere into $SU(2)$,
\be
g(x)= \left( \begin{array}{cc}
             ix_{3}&  x_{1} + i x_{2}\\
             -x_{1} + i x_{2} & -i x_{3}
           \end{array}   \right) \, \label{wmap}
\ee
where the sphere is living inside $ \sRR ^{3}$ with cartesian co-ordinates $(
x_{1}, x_{2},x_{3})$ and $ x_{1}^{2} + x_{2}^{2} +
x_{3}^{2}  = 1$. This map can also be written as
$g(x) = \sum_{k} x_{k}\sigma_{k}$ which defines our conventions for the Pauli
matrices $\sigma_{k}$.

This map is clearly regular (the only non-regular elements of
$SU(2)$ being plus or minus the identity element). It is a smooth map from the
two-sphere to a
two-sphere in $SU(2)$ and is, in fact, the identity map when one considers
$SU(2) \sim S^{3}$ living inside $\sRR ^{4}$ with cartesian co-ordinates $
(x_{1}, x_{2},x_{3},x_{4})$, subject to $x_{1}^{2} + x_{2}^{2} +x_{3}^{2} +
x_{4}^{2} = 1$. We represent elements of $SU(2)$ as
\be
\left( \begin{array}{cc}
            x_{4} + ix_{3}&  x_{1} + i x_{2}\\
             -x_{1} + i x_{2} & x_{4}-i x_{3}
           \end{array}   \right) \, ,
\ee
so that $g$ maps the sphere to itself thought of as the equator of $S^{3}$ ($
x_{4} =0$).

\subs{The Obstruction: Winding Numbers}

To any map $f$ from the two sphere to the two sphere we may
assign an integer, the winding number $n(f)$ of that map.
This is a homotopy invariant
and measures the number of times that the map covers the sphere.
Writing (as above) $f = \sum_{k}f_{k}\sigma_{k}$ with $\sum_{k}(f_{k})^{2}=1$,
an integral representation of its winding number is
\be
n(f) = -\trac{1}{32\pi}\int_{S^{2}} {\rm Tr} f[df,df] \;\;, \label{wno}
\ee
the integral of the the pull-back of the
volume form on the two sphere. This counts the covering by
telling us how many times the volume one picks up. The minus sign appears
in (\ref{wno}) since (in our conventions)
$\Tr \sigma_{k}\sigma_{l} = -2 \d_{kl}$. Clearly for (\ref{wmap}) we
have $n(g)=1$, as can e.g.~be seen by converting the $x_{k}$ to polar
coordinates.

Now suppose that one can smoothly conjugate the map $g$ into a map $t:S^{2}
\rightarrow U(1)$ via some map $h$.
As the space of maps from $S^{2}$ to $SU(2)$ is connected,
$\pi_{2}(SU(2)) = 0$, $g$ is homotopic to $t$. This can be seen by choosing
a homotopy $h_{s}$ between the identity  and $h$ and defining $g_{s}=
h_{s}^{-1}t h_{s}$: then one has $g_{0} = g$ and $g_{1}=t$.
As (\ref{wno}) is a homotopy invariant, one has $n(g) = n(t)$ but,
since $g^{2}=-I$, $t$ is a constant map.
Actually $t$ can be either $\sigma_{3}=$diag$(i,-i)$ or $(-\sigma_{3})$.
We fix on one of
these throughout $S^{2}$ so that $t$ is smooth. This is justified in the
next section. But, as $t$ is constant its winding number is zero, $n(t)=0$,
a contradiction.

More
generally, if $f:S^{2} \rightarrow S^{2} \subset S^{3}$ and one is able
to smoothly conjugate this map to a map into $U(1)$, then one necessarily
has $n(f) =0$. So what we
have learnt is that one may not, in general, smoothly and globally
conjugate into the maximal torus. We will see in the next section that we
can smoothly conjugate into the maximal torus in open neighbourhoods.

\subs{Non-Trivial Torus Bundles}

There is a disadvantage in simply considering the number (\ref{wno}) for it
does not tell us how non-trivial $U(1)$ bundles will arise if we insist, in any
case, on conjugating into $U(1)$, regardless of whether we can do so smoothly
or not.
There is a slight generalisation of the formula (\ref{wno}) which is not only
a homotopy invariant, but for which conjugation (gauge) invariance can be
established directly without any integration by parts.
The advantage of
such a formula is that it allows one to conjugate with arbitrary maps, not
just smooth ones, and so to relate maps which are not homotopic.

Let $A$ be a
connection on the $SU(2)$ product bundle over the sphere. As the bundle is
trivial such an $A$ can be thought of as a Lie algebra valued one-form on
$S^{2}$, $A \in \Omega^{1} (S^{2}, su(2))$. The number we want is
\be
n(f,A) = -\trac{1}{32\pi}\int_{S^{2}} {\rm Tr}f [df,df] -
\trac{1}{2\pi}\int_{S^{2}}  {\rm Tr}
[d(fA)] \, , \label{gno}
\ee
and obviously coincides with (\ref{wno}) when both $f$ and $A$ are smooth.
Furthermore $n(f,A)$ is gauge invariant, i.e.~invariant under simultaneous
transformation of $f$ and $A$,
\be
n(h^{-1}fh, A^{h})=n(f,A) \label{ginv}
\ee
where $A^{h} = h^{-1}Ah + h^{-1}dh$, even for discontinous $h$.
This is seen most readily by rewriting
(\ref{gno}) in manifestly gauge invariant form,
\be
n(f,A) = -\trac{1}{32\pi}\int_{S^{2}} {\rm Tr}f [d_{A}f,d_{A}f ] -
\trac{1}{2\pi}\int_{S^{2}}  {\rm Tr}[fF_{A}] \, , \label{cno}
\ee
with $d_{A}f=df +[A,f]$ and $F_{A} = dA + \trac{1}{2}[A,A]$.

Let us now choose $h$ so that it conjugates our favourite map $g$ into $U(1)$,
say $ g = h^{-1}\sigma_{3} h$. Using (\ref{ginv}) we find
\be
n(g,A)=1= -\trac{1}{2\pi}\int_{S^{2}} \Tr \sigma_{3} d(A^{h^{-1}})\;\;.
\label{chno}
\ee
In particular, if we introduce the Abelian gauge field $a=-\Tr\sigma_{3}
A^{h^{-1}}$ (we will see in section 5 that this is consistent with the
gauge transformations of $a$ and the global geometry of the problem)
we obtain
\be
n(g,A) = 1 = \trac{1}{2\pi}\int_{S^{2}} da\;\;.
\ee
We now see the price of conjugating into the torus. The first Chern class of
the $U(1)$ component of the gauge field $A^{h^{-1}}$ is equal to the winding
number of the original map! We have picked up the sought for non-trivial
torus bundles.

If one chooses to conjugate $g$ into $(-\sigma_{3})$ instead by replacing
$h$ by $nh$ for a suitable $n\in N(\bT)$, the expression (\ref{chno}) remains
invariant as
\be
\Tr(-\sigma_{3}) A^{(nh)^{-1}} = \Tr \sigma_{3} A^{h^{-1}}\;\;.
\ee
Thus the $\bT$-bundle which emerges is independent of the choice of $t$
and is hence canonically associated with the original map $g$.
In this case it is just the pull-back of the $U(1)$-bundle
$SU(2)\ra SU(2)/U(1)\sim S^{2}$ via $g$ and this turns out to be more or
less what happens in general. As both $g$ and its diagonalization
$\pm\sigma_{3}$ may just as well be regarded as Lie algebra valued maps,
this example establishes that obstructions to diagonalization
will also arise in the
(seemingly topologically trivial) case of Lie algebra valued maps.

It is possible to generalise both (\ref{wno}) and (\ref{cno}) to other
manifolds and to other groups (we will do so further along).
In due course we will also tie these up with some general results on the
classification of torus bundles over surfaces.

\subs{Example 3: A Map from $S^{1}$ to $SO(3)$}

While we have seen in example 1 that non-regularity is one obstruction
to finding a globally well-defined smooth diagonalization $t$, even for
regular $g$ an obstruction to finding such a $t$ may arise. We will
establish in section 4 that this can only happen when neither
$\bG$ nor $M$ is simply connected. The raison d'\^etre of this obstruction
is the fact that diagonalization involves lifting a map into $\bG_{r}$
to a map into $\gt\times\bT_{r}$ which may not be possible if the fibration
(\ref{w1}) is non-trivial. Here we illustrate this obstruction
by a map from $S^{1}$ into $SO(3)_{r}$ (cf.~the remarks at the end of section
2).

Consider first of all the following path in $SU(2)_{r}$,
\be
\tg(x) =
\trac{1}{\sqrt{2}}\mat{e^{ix/2}}{ie^{-ix/2}}{ie^{ix/2}}{e^{-ix/2}}\;\;.
\ee
As $\tg(2\pi)=-\tg(0)$, $\tg$ will project to a non-contractible loop
$g\equiv{\rm Ad}(\tg)\in\map{S^{1}}{SO(3)_{r}}$.
Explicitly, this $g$, satisfying
\be
\tg^{-1}\sigma_{k}\tg = g_{kl}\sigma_{l}
\ee
and $g(2\pi)=g(0)$, is given by
\be
g(x) = \som{0}{0}{1}{\sin x}{\cos x}{0}{-\cos x}{\sin x}{0}\;\;.
\ee
There is no obstruction to diagonalizing $\tg$, $\tg = \th^{-1}\tti\th$
and there are two solutions $\tti_{\pm}$ differing by a Weyl transformation
(exchange of the diagonal entries). It can be checked that
 $\tti_{\pm}(2\pi)$ differs from $\tti_{\pm}(0)$ not only
by a sign but also by a Weyl transformation,
\be
\tti_{\pm}(0)=\trac{1}{\sqrt{2}}\mat{1\pm i}{0}{0}{1\mp i} =
-\tti_{\mp}(2\pi)\;\;.
\ee
Hence $\tti$ will not project to a closed loop in $SO(3)$ and the
diagonalization $t$ of $g$ will necessarily be discontinuous (non-periodic),
as can also be checked directly. Choosing the torus $SO(2)
\ss SO(3)$ to consist of elements of the form
\be
\som{\cos y}{-\sin y}{0}{\sin y}{\cos y}{0}{0}{0}{1}\;\;,
\ee
with the Weyl group acting as $y\ra -y$, one finds that
\be
t(0) = \som{0}{-1}{0}{1}{0}{0}{0}{0}{1}\;\;,
\ee
while
\be
t(2\pi)  = \som{0}{1}{0}{-1}{0}{0}{0}{0}{1}\;\;.
\ee
Hence the periodic regular map $g$ cannot be diagonalized to a periodic
map $t$ and, regarded as map from $S^{1}$ into $SO(2)_{r}$, $t$ will only
be smooth locally.

This concludes our visit to the
zoo of obstructions and we now turn to establishing
that at least locally regular maps can always be smoothly conjugated into the
maximal torus.

\section{Local Conjugation to the Maximal Torus and Global Obstructions}

Let $P_{G}$ be a principal $\bG$ bundle over a smooth connected manifold
$M$ and denote by
${\rm Ad}P_{G}$ the group bundle associated to $P_{G}$ via the adjoint action
of $\bG$ on itself and let $g$ be a section of ${\rm Ad}P_{G}$. Locally,
i.e.~over a trivializing open neighbourhood $U\ss M$, $g$ can be regarded
as a $\bG$ valued map transforming in the adjoint representation. We will from
now on take $U$ to be a contractible open set. We assume
that $g$ is regular, i.e.~takes values in $\bG_{r}\ss \bG$. Note that

\noindent a)
the regularity of $g$ is independent of the chosen local trivializations of
$P_{G}$ (as $\bG_{r}$ is preserved by the adjoint action), and

\noindent b) regular $g$ always exist when $P_{G}$ is trivial (for
more comments on the existence of regular sections see section 6).

Being able to locally conjugate smoothly into the maximal
torus is the statement that we can find smooth maps $h_{U}\in\ug$
and $t_{U}\in\ut$ such that the restriction $g_{U}$ of $g$ to $U$
can be written as $g_{U} = h_{U}^{-1}t_{U} h_{U}$. In other words, we
are looking for a (local) lift of the map $g\in\mgr$ to a map
$(h,t)\in\mg\times\mtr$. We will establish the
existence of this lift in a two-step procedure indicated in diagram
(4.1).
\be
\bfig
\putsquare<1`-1`1`1;1000`1000>(500,0)%
[\bG\times\bT_{r}`\bG/\bT\times\bT_{r}`M`\bG_{r};p\times 1`(h,t)`q`g]
\putmorphism(1500,1000)(-1,-1)[``(f,t)]{1000}{-1}{l}
\efig
\ee
In the first step we lift $g$ along the diagonal, i.e.~we
construct a pair $(f,t)$, where $f\in\mgt$,
which projects down to $g$ via the projection $q$ introduced in
(\ref{w1}). The obstruction to doing this globally is related to the
possibility of having non-trivial $W$ bundles on $M$ (as in examples 1 and 3
of the previous section) but only arises if neither $\bG$ nor $M$ is
simply connected.

In the second step, dealing with the upper triangle, we will lift $f$ locally
to $\mg$, and the obstruction to doing this globally is given
by non-trivial $\bT$ bundles on $M$ (as in example 2).

\subs{The First Lifting-Problem: $W$-Bundles}

We begin by recalling that the conjugation map $q:\bG/\bT \times \bT_{r}
\rightarrow \bG_{r}$, given by $([h],t) \mapsto h^{-1}th$, is a
smooth $|W|$-fold covering of $\bG_{r}$ so that
$\bG/\bT \times
\bT_{r}$ is the total space of a principal fibre bundle over
$\bG_{r}$ with fibre and structure group
$W$ and projection $q$. Given the map $g$ into $\bG_{r}$, the base
space of this bundle, we would like
to lift this to a map into the total space, i.e.~we want to find a pair
$(f,t)\in\mgt\times\mtr$ such that diagram (4.2) commutes.
\be
\dtriangle<-1`1`1;700>[\bG/\bT \times \bT_{r}`M`\bG_{r};(f,t)`q`g]
\ee
That such a map indeed exists locally is a consequence of the following
fundamental result on the lifting of maps (see e.g.~\cite{hu} for this
and most of the other topological results used in this paper):
If $P$ is a (smooth)
principal fiber bundle with
base space $B$ and $f$ is a (smooth) map from a manifold $X$ to $B$ then
$f$ can be lifted to a (smooth) map into $P$ if and only if the pull-back
bundle
$f^{*}P$ over $X$ is trivial. It is indeed easy to see that there is a
direct correspondence
between lifts of $f$ and trivializing sections of $f^{*}P$.

The first implication of this result is that locally, i.e.~over some
contractible open set $U\ss M$, the desired lift can always be found
as the pull-back bundle will certainly be trivializable over $U$.

However, in certain cases we can sharpen this statement to establish
the existence of a global lift. Consider e.g.~the case when $\bG$ is
simply connected.
As the principal $W$-bundle $\bG/\bT\times \bT_{r}\ra\bG_{r}$ is then
trivial, so is its pull-back to $M$ via any map $g\in\mgr$.
Hence a lift $(f,t)$ making the above diagram commute exists globally on
$M$. There is an
obvious $|W|$-fold ambiguity in the choice of such a lift.

Even if $\bG$ is not simply connected but $M$ is, the pull-back bundle
is necessarily trivial over $M$ (otherwise it would be a non-trivial
covering of $M$) and again a lift $(f,t)$ will exist globally.

Finally, there is a class of maps for which the $W$-obstruction does not
arise regardless of what $M$ and $\bG$ are. This class consists of those
maps $g$ which are conjugate to a constant map $t$ into $\bT$. We will have
more to say about these maps and why they are interesting in section 7.

\subs{The Second Lifting Problem: $\bT$-Bundles}

It remains to lift the $\bG/\bT$ valued map $f$ to $\bG$. Thus
we are looking for a $h\in\mg$
making the following diagram commute (with the replacement of
$M$ by $U$ if only the local existence of $(f,t)$ could be established):
\be
\dtriangle<-1`1`1;700>[\bG`M`\bG/\bT;h`p`f]
\ee
Here $p$ is the projection of the principal fibration
$p:\bG\ra\bG/\bT$.
By construction this map will then satisfy $g= h^{-1}th$. However, by the
same result on the lifting of maps quoted above there will be an
obstruction to finding such an $h$ globally.
As $\bG$ can be regarded as the total space of
a principal $\bT$-bundle over $\bG/\bT$, the same reasoning as above leads us
to conclude that such a lift exists iff $f^{*}\bG$ is a trivial(izable) $\bT$
bundle over $M$. Whether or not this is the case will depend on the interplay
between the homotopy class of $f$ and the classification of torus bundles on
$M$. However, if we restrict $f$ to $U\ss M$, then a lift $h_{U}$
of $f$ over $U$ will always
exist as the pull-back bundle is certainly trivializable
over the contractible set $U$. The upshot of this is that, for a regular map
$g$
we can always locally find smooth $\bG$-valued functions $h_{U}$ such that
$h_{U}g_{U}h_{U}^{-1}$ takes values in $\bT_{r}$.

We summarize the results about the possibility to conjugate a map locally
into a maximal torus in

\noindent{\bf Proposition 1:} Let $\bG$ be a compact Lie group, $\bT$ a
maximal torus, $M$ a smooth
manifold, $U \ss M$ a contractible open set in $M$, $P_{G}$ a principal
$\bG$ bundle over $M$ and $g$ a section of ${\rm Ad}P_{G}$.
If $g|_{U}\equiv g_{U}$ is regular, then it can be smoothly
conjugated into $\bT$. In other words, under these circumstances
there exist smooth functions $t_{U}\in\utr$ and $h_{U}\in\ug$ such that
$g_{U} = h_{U}^{-1}t_{U}h_{U}$.

Of course, we already know a little bit more than that, for instance that
under certain conditions the diagonalized map $t$ will exist globally.
We can also be
more precise about the obstruction occurring in the second lifting problem,
as torus bundles
are classified by $H^{2}(M,\ZZ^{r})$, where $r=\dim\bT$ is the rank of $\bG$.
We have therefore established the following results concerning global
obstructions to conjugating a map $g:M\ra \bG_{r}$ into the torus:

\noindent{\bf Proposition 2:} Let $g:M\ra \bG_{r}$ be a smooth regular map.
Then a smooth map $t:M\ra \bT_{r}$ satisfying $g=h^{-1}th$ for some (not
necessarily smooth) map $h:M\ra\bG$ exists globally if
$g^{*}(\gt\times\bT_{r})$ is the total space of a trivial $W$-bundle over
$M$. If, furthermore, $f^{*}(\bG)$ (where
$f$ is the $\bG/\bT$-part of the lift of $g$) is a trivial $\bT$ bundle over
$M$, then $h$ can be chosen to be smooth globally.

\noindent{\bf Corollary 1:} If either $M$ or $\bG$ is simply connected, a
smooth diagonalization $t\in\mtr$ of a regular $g$ will exist globally.
If, moreover, $H^{2}(M,\ZZ)=0$, then a smooth regular map $g$
can be smoothly conjugated into a maximal torus, i.e.~there exists a smooth
function $h\in\mg$ such that $g = h^{-1}t h$.

As loop groups are a particularly interesting and well studied class of
spaces of group valued maps \cite{ps},
we also mention seperately the following immediate consequence of the above
considerations:

\noindent{\bf Corollary 2:} If $\bG$ is simply connected, every regular
element of the group ${\bf L}\bG$ of smooth loops in $\bG$ can be smoothly
diagonalized.

Examples 1 and 3 of section 3 show that both regularity and simple
connectivity are necessary conditions. What we have shown is that they
are also sufficient.

At least when $\bG$ is simply connected, there is a slightly more canonical
way of describing the results obtained in Proposition 2, one which does not
depend on the (arbitrary) choice of a maximal torus $\bT$ of $\bG$. We first
observe that over $\bG_{r}$ there is a natural torus bundle $P_{C}$ (the
centralizer bundle) with total space
\be
P_{C}=\{(g_{r},g)\in\bG_{r}\times\bG:g\in C(g_{r})\}
\ee
and projection $(g_{r},g)\mapsto g_{r}$. For any map $g\in\mgr$ this bundle
can be pulled back to a torus bundle $g^{*}P_{C}$ over $M$ and it is the
possible non-triviality of this bundle which is the obstruction to finding
a globally smooth $h$ accomplishing the diagonalization. To make contact
with the previous construction, we note that under the isomorphism
$q: \gt\times\bP_{r}\ra\bG_{r}$ the bundle $P_{C}$ pulls back to the
$\bT$-bundle $\bG\times\bP_{r}\ra\gt\times\bP_{r}$, while the lift $(f,t)$
in diagram (4.2) can be written as $(f,t) = q^{-1}\circ g$. This is
illustrated in the diagram below.
\be
\bfig
\putsquare<1`1`1`1;700`700>(300,0)%
[g^{*}P_{C}`P_{C}`M`\bG_{r};\hat{g}```g]
\putsquare<-1`0`1`-1;700`700>(1050,0)%
[`\bG\times\bP_{r}``\gt\times\bP_{r};\hat{q}``p\times 1`q]
\efig
\ee

\subs{Conjugation of $\lg$-valued Maps into the Cartan Subalgebra}

The question of diagonalizability of Lie algebra valued maps (the case
of interest in e.g.~Yang-Mills or Chern-Simons theory) can be addressed
in complete analogy with the analysis for group valued maps performed above.
It will turn out that the only substantial difference between the two
is that the first obstruction
(non-trivial $W$-bundles) does not arise. That the second obstruction,
related to non-trivial torus bundles, persists can already be read off
from example 2 of section 3 as the map $g = \sum_{k}x_{k}\sigma_{k}$
and its diagonalization $t=\pm\sigma_{3}$ considered there can equally
well be regarded as Lie algebra valued maps.

Let us denote by $\lg$ and $\lt$ (a Cartan subalgebra of $\lg$)
the Lie algebras of $\bG$ and $\bT$ respectively and by $\lg_{r}$ and
$\lt_{r}$ their regular elements. As in (\ref{w1}) there is a smooth
$|W|$-fold covering
\bea
&&q':\bG/\bT\times\lt_{r}\ra \lg_{r}\;\;,\nonumber\\
&&q'([h],\tau) = h^{-1}\tau h\;\;.
\eea
However, $\lg$ is a vector space and hence simply connected. As a consequence
$\lg_{r}$ is simply connected as well. Therefore this $W$-bundle
is necessarily trivial
and the first lifting problem can always be solved globally on $M$. This
establishes the global existence of a lift $(f,\tau)$ of a smooth map
$\phi\in\map{M}{\lg_{r}}$ to
$\gt\times\lt_{r}$. In particular, a smooth global diagonalization
$\tau\in\map{M}{\lt_{r}}$ of $\phi$ always exists.

The second lifting problem depends only on
the $\gt$-part $f$ of the lift and is identical with that for group valued
maps.
Therefore the situation concerning Lie algebra valued maps is the following:

\noindent{\bf Proposition 3:} Let $\phi\in\map{M}{\lg_{r}}$
be a smooth regular map
into the Lie algebra $\lg$ of a compact Lie group. Then a smooth
diagonalization $\tau\in\map{M}{\lt_{r}}$ exists globally.
If  $f^{*}\bG$ is the
total space of a trivial principal $\bT$-bundle over $M$ then there exists
a smooth functions $h\in\mg$ such that $\phi = h^{-1}\tau h$ globally.

\noindent{\bf Corollary 3:}
If $H^{2}(M,\ZZ)=0$, any $\phi\in\map{M}{\lg_{r}}$ can be
smoothly diagonalized.

\section{Global Conjugation and Non-Trivial Torus Bundles}

Having demonstrated that we are able to conjugate into the maximal Torus
smoothly in open contractible neighbourhoods we turn to more global questions.
In particular we want to investigate what happens when we perform the
diagonalizations patch-wise and then try to glue the resulting data together
globally. This has a three-fold purpose. Firstly, it allows us to understand
the obstructions we encountered above in a more down-to-earth way by
constructing explicitly the non-liftable maps into $\bG/\bT$ and the transition
functions of non-trivial $\bT$-bundles in terms of the local data
$(h_{U},t_{U})$. Secondly it will allow us to establish a partial converse to
the above results in that we are e.g.~able to show that under certain
circumstances any non-trivial $\bT$ bundle will appear as the obstruction for
some map. And thirdly it is the most convenient language to analyse what
happens to gauge fields when one insists on diagonalizing globally by a
collection of locally defined $h$'s. Unless stated otherwise, $\bG$ will
be assumed to be simply connected in this section.

\subs{Glueing the Local Data}

Let again $g\in\mgr$ be a regular smooth map and
let us choose a covering of $M$ by open contractible sets $\{\ua\}$.
By Proposition 1 we have, on each $\ua$, smooth
functions $(\ha,\ta)$ such that $\ga = \ha^{-1}\ta\ha$ where $\ga = g|_{\ua}$.
Actually, we already know that the $\ta$'s glue together to a global map
(we are assuming that $\pi_{1}(\bG)=0$), but we will rederive this result
below in a different way, one which gives more insight
into why the condition of regularity is so crucial.

As $g$ is globally defined, on the overlaps $\uab$ of patches we must have
\be
h_{\a}^{-1}t_{\a}h_{\a}  =  h_{\b}^{-1}t_{\b}h_{\b} \;\;.
\ee
Put another way, the $\ta$'s in different patches are related by
\be
t_{\a} = h_{\a \b}t_{\b}h_{\a \b}^{-1}\;\;, \label{overlap}
\ee
where
\be
h_{\a \b} =h_{\a}h_{\b}^{-1} \;\;. \label{tcocycle}
\ee
The functions $\hab$, a priori taking values in $\bG$, can be regarded as
the transition functions of a trivial $\bG$ bundle on $M$ (that they
satisfy the cocylce condition $\hab\hbc = \hac$ is clear from the
definition (\ref{tcocycle})). The $\ta$ thus transform as sections of the
(trivial) adjoint bundle ${\rm Ad}P_{G}$ of $P_{G}\sim M\times\bG$.

\subs{Restrictions of the Structure Group and Torus Bundles}

The first thing we will show is that regularity of the $\ta$ implies
that the transition functions $\hab$ take values in the normalizer $N(\bT)$.
To see that, note that pointwise (\ref{overlap}) implies that $\ta$ is
contained not only in $\bT$ but also in the maximal torus
$\hab\bT\hab^{-1}$. By regularity of $\ta$ this implies that
$\bT=\hab\bT\hab^{-1}$.
{}From this we conclude that the $h_{\a \b}$ take values in the normalizer of
$\bT$ in $\bG$,
\be
\hab : U_{\a} \cap U_{\b} \rightarrow N(\bT)\;\;,
\ee
so that a choice of $\ha$'s diagonalizing $g$ gives rise to a
restriction of the structure group of the (trivial) principal $\bG$-bundle
to $N(\bT)$ (for the precise definition of the restriction of structure
groups see section 6). Then (\ref{overlap}) means that on overlaps $\uab$
the $\ta$'s are related by Weyl transformations.

So far the condition
$\pi_{1}(\bG)=0$ has not entered, in agreement with the results of section
4 which allow one to read off that the ambiguity in patching together the
local solutions $\ta$ can always be reduced to a $W$ ambiguity.

If, however, $\bG$ is simply connected, by choosing the $\ha$'s
appropriately, the $\hab$'s can be arranged to take
values in $\bT$ and hence the structure group will be reduced to $\bT$. For
$\bG=SU(n)$ this can be done by adopting some particular ordering
prescription for the diagonal entries, e.g.~according to size. As no two
eigenvalues are the same for regular elements, this eliminates the Weyl
ambiguity (permutation of the diagonal entries). And in general
this is achieved in the following way. Let us choose a fundamental domain $D$
for the action of the Weyl group $W$ on $\bT_{r}$, i.e.~some identification
of $\bT_{r}/W$ with a connected component $D\sim\bP_{r}$ of $\bT_{r}$ (this is
where the assumption of simple connectivity of $\bG$ enters in a crucial way).
Let us now, using the
ambiguity $\ha\ra n_{\a}\ha$, $n_{\a}\in\map{\ua}{N(\bT)}$,
choose the $\ha$'s in such a way that the $\ta$'s take values in $D$.
In the language of the previous
section this choice corresponds to a choice of lift in diagram (4.2).
Then (\ref{overlap}),
read as $\ta = w_{\a\b}(\tb)$, $w_{\a\b}\in W$, implies $w_{\a\b}=1$ as
$\ta$ and $\tb$ take values in $D$ while $D\cap w(D)=\emptyset$ unless $w=1$.
But this implies  that
\be
\ta=\tb \;\;\;\;{\rm on}\;\;\;\;\uab \;\;,\label{tr1}
\ee
so that there exists a globally defined smooth $t=\{\ta\}$
(as we had already seen in the previous section). It also implies
that the $\hab$ are $\bT$-valued functions on the overlaps $\uab$,
\be
\hab\equiv\ha\hb^{-1}:\uab\ra\bT\label{tr2}\;\;.
\ee
Thus the $\hab$ define a (possibly non-trivial) torus bundle on $M$ which is
however trivial when regarded as a principal $\bG$ bundle (since $\hab =
\ha\hb^{-1}$). Using the trivial identity $\ha = \hab \hb$ we also see that
we can interpret the $\ha$'s as (local trivializing)
sections of the principal $\bT$ bundle
$P_{T}$ with fibre $\bT\ha(x)=\{t\ha(x), t\in\bT\}$
above the point $x\in M$, establishing once
again that the $\ha$'s can be defined globally iff the bundle $P_{T}$ is
trivial.

Thus, given a regular smooth map $g$ we obtain, upon choice of a fundamental
domain $D$, a smooth $\bT_{r}$-valued map $t$ and a principal $\bT$ bundle
$P_{T}$ characterized by the transition functions $\hab$. We now want to
establish a converse to this result.

Let us assume that $M$ and $\bG$ are
such that all $\bG$-bundles on $M$ are trivial.
Given some $\bT$ bundle $P_{T}$ and a regular smooth map $t$,
one obtains a map $g$ as follows. As every $\bT$ bundle is, in
particular, a $\bG$ bundle ($\bT\ss\bG$), regarded as a $\bG$ bundle
$P_{T}$ is necessarily trivial. This means that the transition functions
$h_{\a \b}$ on $U_{\a} \cap U_{\b}$ that define $P_{T}$ can be
expressed as $h_{\a \b} = h_{\a}h_{\b}^{-1}$ where the $h_{\a}$ are
$\bG$-valued maps on the respective patches $U_{\a}$. Armed with this data one
defines on each patch $U_{\a}$
\be
\ga = h_{\a}^{-1}\ta h_{\a} \label{conv1}
\ee
where $\ta = t|_{\ua}$. These $\ga$ patch together to define a globally
well-defined smooth $\bG_{r}$-valued map since on overlaps one has
\bea
\ga &=& h_{\a}^{-1}\ta h_{\a} \nonumber \\
& = & h_{\b}^{-1}h_{\a \b}^{-1}\ta h_{\a \b} h_{\b} \nonumber \\
& = & h_{\b}^{-1}\tb h_{\b} = \gb\;\;.   \label{conv2}
\eea
By construction, this map $g$ will give rise to the transition functions of
$P_{T}$ upon diagonalization. In particular, therefore, in the case at hand
every  isomorphism class of torus bundles will appear as the obstruction to
global diagonalizability for some regular map $g$.
We will see below that this can also be understood
directly in terms of classifying maps and universal bundles.

\subs{Maps into $\bG/\bT$}

We have seen in section 4 that, in addition to a smooth $\bT_{r}$ valued
map $t$, a regular $g$ also gives rise to a map $f\in\mgt$ governing
the obstruction to conjugating $g$ into $\bT$ smoothly. In terms of the
local data $\ha$ associated with $g$ and $t$, i.e.~chosen to be
compatible with a
fixed fundamental domain $D$, these maps can be constructed in the following
way.

We realize $\bG/\bT$ as the coadjoint orbit ${\cal O}_{\mu}$
through a regular element
$\mu\in\lt^{*}$ ($\lt$ denoting the Lie algebra of $\bT$)
so that e.g.~the principal fibration $p:\bG\ra\bG/\bT$ can
be written as $p(g)=g^{-1}\mu g$ (see the discussion following
(\ref{coado}) below for more on coadjoint orbits).
We then define local $\bG/\bT$-valued maps $f_{\a}$ by
\bea
&&f_{\a}:\ua\ra{\cal O}_{\mu}\nonumber\\
&&f_{\a}=h_{\a}^{-1}\mu h_{\a}\;\;.\label{tr4}
\eea
As upon a choice of $D$ the
$\ha$'s are unique up to left-multiplication by $\bT$-valued maps, the
$f_{\a}$'s are well-defined and independent of which $\ha$'s one chooses.
On overlaps $\uab$ one finds that
\bea
f_{\b}&=& \hb^{-1}\mu \hb\nonumber\\
      &=& \ha^{-1}(\ha\hb^{-1}) \mu (\hb \ha^{-1}) \ha \nonumber\\
      &=& \ha^{-1}\hab\mu\hab^{-1}\ha\nonumber \\
      &=& \ha^{-1}\mu\ha = f_{\a}\;\;, \label{tr3}
\eea
since the transition functions $\hab$ are $\bT$-valued and act trivially on
$\mu$. Thus the $f_{\a}$ define a globally well-defined map $f\in\mgt$
whose local lifts to $\bG$ are given by the $\ha$, as in diagram (4.3).
It is clear that $f$ is conjugate to the constant map $\mu$ if a diagonalizing
$h$ exists globally.

We shall see below how, for simplicity in the case that $M$ is two-dimensional,
the winding numbers of $f$ are related to the Chern classes of the
corresponding torus bundle over $M$.

\subs{Relation between Connections on $\bG$ and $\bT$ Bundles}

We consider again the case of regular maps $g\in\mgr$, i.e.~sections of
the adjoint bundle associated to the trivial bundle $P_{G}\sim M\times \bG$.
Via a choice of trivialization, connections on $P_{G}$ can be identified with
Lie-algebra valued one-forms on $M$. Gauge transformations (vertical
automorphisms of $P_{G}$) can be identified with sections $h$ of
${\rm Ad}P_{G}$ and the induced action of $h$ on $g$ is given by
$g\ra hgh^{-1}$. Thus, when considering connections on $P_{G}$, such a
transformation has to be accompanied by a gauge transformation on the gauge
fields, $A\ra hAh^{-1}+h dh^{-1}$. We now look at what happens
to gauge fields when we gauge transform them patch-wise using the diagonalizing
maps $\ha$. Let the connection
obtained in this way on an open set $U_{\a}$ be denoted by $A_{\a}$. On the
overlap $U_{\a} \cap U_{\b}$ one has
\be
A_{\a} = h_{\a \b} A_{\b} h_{\a \b}^{-1} + h_{\a \b} d h_{\a \b}^{-1} \,.
\ee
Decomposing the Lie-algebra $\lg$ as $\lg = \lt \oplus \lk$ and
correspondingly the gauge field as $A_{\a}=A_{\a}^{\lt}+A_{\a}^{\lk}$,
one finds that
\bea
A_{\a}^{\lt} &=& A_{\b}^{\lt} + \hab d\hab^{-1}\;\;,\nonumber\\
A_{\a}^{\lk} &=& \hab A_{\b}^{\lk} \hab^{-1}\;\;,     \label{connt}
\eea
as the $\hab$ take values in $\bT$. Thus only the $\lt$-component of the new
gauge field $A^{h^{-1}} = \{A_{\a}\}$ transforms inhomogeneously and defines
a connection on the torus bundle $P_{T}$ determined by the transition functions
$\hab$. The $\lk$-component, on the other hand, transforms as a one-form with
values in sections of the
bundle $P_{T}\times_{\bT}\lk$ associated to $P_{T}$ via the adjoint action of
$\bT$ on $\lk$. The very same conclusions can be reached if one starts off
with a connection on a non-trivial $\bG$-bundle admitting a restriction to
$\bT$ and then proceeds with the local analysis as in section 6.

\subs{Classification of Torus Bundles}

We would now like to bring some of the threads together which have appeared
in this and the previous section. We have seen that, associated with a
regular map $g\in\mgr$ and a connection $A$ on $P_{G}\sim M\times \bG$
we have the following data:
\begin{itemize}
\item a smooth map $t\in\mtr$, unique up to $W$-transformations;
\item a corresponding collection of maps $\ha\in\uag$, unique up to
      multiplication by $\bT$-valued functions on the left;
\item a principal $\bT$ bundle $P_{T}$, determined by the transition functions
      $\hab = \ha\hb^{-1}$;
\item a smooth map $f\in\mgt$, uniquely determined by the choice of $t$;
\item a connection $A^{\lt}=\{A^{\lt}_{\a}\}$ on $P_{T}$;
\item a one-form $\{A^{\lk}_{\a}\}$ with values in the sections of the
      associated bundle $P_{T}\times_{\bT}\lk$.
\end{itemize}
On the basis of the arguments presented in section 4 one would expect there
to be a close relation between the topological type (Chern classes) of $P_{T}$
and the homotopy class (winding numbers) of $f$. To make this relation as
explicit as possible, we consider in the following the case of two-dimensional
orientable manifolds $M=\Sigma$ (and, as before, simply connected groups).

In order to proceed it will be helpful to make use of the notions of universal
bundles and classifying spaces \cite{hu}.
By definition, a universal $\bK$ bundle,
$\bK$ some (compact) group, is a principal $\bK$ bundle with contractible
total space $E\bK$.
It can be shown that isomorphism classes of principal
$\bK$ bundles on $M$ are in one-to-one correspondence with homotopy classes
of based maps from $M$ to $B\bK = E\bK/\bK$, the base space of the universal
bundle. For this reason, $B\bK$ is called the classifying space.
The correspondence is given by pull-back,
that is, every principal $\bK$ bundle over $M$ can be realized as $f^{*}E\bK$
for some $f:M\ra B\bK$ and two bundles $f_{1}^{*}E\bK$ and $f^{*}_{2}E\bK$ are
isomorphic if and only if $f_{1}$ and $f_{2}$ are homotopic.

Usually, $E\bK$ is
infinite-dimensional (e.g. $EU(1)= S^{\infty}$ with $BU(1)=\CC\PP^{\infty}$),
but if one is only interested in classifying bundles in (or up to) a given
dimension $n$ it is sufficient to consider a bundle $E\bK^{(n)}\ra B\bK^{(n)}$
which is $n$-universal, i.e.~for which $\pi_{0}(E\bK^{(n)})=\pi_{1}(E\bK^{(n)})
=\ldots =\pi_{n}(E\bK^{(n)}) = 0$. $n$-universal bundles and $n$-classifying
spaces can be chosen to be finite dimensional (and are typically
Stiefel-bundles over Grassmannians).

Thus, in order to classify torus bundles over a surface $\Sigma$,
we need a $2$-universal bundle $E\bT^{(2)}$. If $\bG$ is simply-connected,
then the bundle $\bG\ra\bG/\bT$ with $B\bT^{(2)}=\bG/\bT$ precisely
satisfies this requirement as one has
$\pi_{0}(\bG) = \pi_{1}(\bG) = \pi_{2}(\bG) = 0$ in this case.
This means that isomorphism classes of torus bundles
are in one to one correspondence with homotopy classes of maps from
$\Sigma$ to $\bG/\bT$. As $\gt$ is simply connected, $\pi_{1}(\gt)=0$,
up to the two-skeleton it can be regarded as an Eilenberg - MacLane
space $K(\pi,2)$, where $\pi = \pi_{2}(\gt)$ and one therefore has
\be
[\Sigma,B\bT]_{*} \sim H^{2}(\Sigma,\pi_{2}(\bG/\bT))\sim \pi_{2}(\bG/\bT)
\sim \ZZ^{r}\;\;.
\ee
Hence torus bundles on $\Sigma$ are classified by an $r$-tuple
of integers measuring the winding around non-contractible two-spheres in
$\bG/\bT$.

In particular, therefore, with each regular map $g\in\sgr$ there
is associated an $r$-tuple of integers. Furthermore, it is clear (and can
also be read off from the integral representations of the winding numbers
given below) that these integers do not change under regular homotopy,
i.e.~under a homotpy $g_{s}$, $s\in [0,1]$ between $g_{0}$ and $g_{1}$
where $g_{s}$ is regular for all $s$. Now, all maps from $\Sigma$ to $\bG$
are homotopic as
\be
\pi_{0}(\sg) = \pi_{2}(\bG) = 0\;\;.
\ee
In particular, therefore, all maps into $\bG_{r}$ can be homotoped into each
other (by possibly non-regular homotopies).
But, as a consequence of (\ref{p2gr}), the
space of regular maps consists
of a $\ZZ^{r}$'s worth of disconnected components,
\be
\pi_{0}(\sgr) = \pi_{2}(\bG_{r}) = \ZZ^{r}\;\;.
\ee
Because the proof of (\ref{p2gr}) hinges on the fact that $\pi_{2}(\gt)$
and $\pi_{2}(\bG_{r})$ are equal under the isomorphism provided by the
conjugation map $q$, it also follows that two maps $g_{0}$ and $g_{1}$
are regularly homotopic if and only if they give rise to homotopic
maps $f_{0}$ and $f_{1}$ into $\gt$ (i.e.~to maps with the same winding
numbers). Thus we have shown

\noindent{\bf Proposition 4:} Let $\bG$ be a simply connected compact
Lie group and $\Sigma$ a two-dimensional manifold. Then $\pi_{0}(\sgr)
= \ZZ^{r}$ and two regular maps are regularly
homotopic if and only if their lifts to $\gt$ are homotopic in $\sgt$.

\subs{Coadjoint Orbits and Symplectic Forms}

The winding numbers of the maps into $\gt$
can be given an integral representation
which we shall use to relate them to the Chern classes of torus bundles,
as in the discussion in example 2 of section 3.  They are
best understood in terms of the Kirilov-Kostant symplectic forms on $\gt$
thought of as a coadjoint orbit of $\bG$, i.e.~in terms of volume forms
on the non-contractible two-spheres in $\gt$. We thus make a short excursion
into the symplectic geometry of coadjoint orbits.

For $\mu$ a regular element of $\lt\sim\lt^{*}$ let ${\cal O}_{\mu}$
be the coadjoint orbit through $\mu$,
\be
{\cal O}_{\mu} = \{g^{-1}\mu g:\;g\in\bG\}\;\;.\label{coado}
\ee
Because $\mu$ is regular, the stabilizer of $\mu$ is isomorphic to $\bT$ and
one has ${\cal O}_{\mu}\sim \gt$. The coadjoint orbit comes
equipped with a natural symplectic form (Kirilov-Kostant form) which is defined
as follows. The infinitesimal version of the $\bG$-action on ${\cal O}_{\mu}$
is $\d_{X}\mu = [\mu,X]$ with $X\in\lg$ defined mod $\lt$. As the $\bG$ action
is transitive, the tangent space to ${\cal O}_{\mu}$ at $\mu$ is spanned by
tangent vectors of this form. The symplectic form at $\mu$ is defined by
\be
\omega_{\mu}(\d_{X}\mu,\d_{Y}\mu) = \Tr \mu [X,Y] \label{kkf}
\ee
and extended to all of ${\cal O}_{\mu}$ by the $\bG$-action. It is easily
verified that the right hand side of (\ref{kkf})
depends on $X$ and $Y$ only modulo $\lt$ and
that it defines a closed non-degenerate two-form on ${\cal O}_{\mu}$,
i.e.~a symplectic form. Varying $\mu$ in $\lt_{r}$ one obtains an
$r$-parameter family of symplectic forms on $\gt$ and for certain values
of $\mu$ these symplectic forms are integral. Let us assume that $\Tr$
is normalized in such a way that the $r$ symplectic forms
$\omega^{k}, k=1,\ldots,r$ for $\mu = \a^{k}$ a simple root
are the generators of $H^{2}(\gt,2\pi\ZZ)\sim 2\pi\ZZ^{r}$.

One can then assign $r$
homotopy invariant integers $n^{k}(f)$ to any map $f\in\sgt$ by
\be
n^{k}(f) = \trac{1}{2\pi}\int_{\Sigma} f^{*} (\omega^{k})  \label{c}
\ee
which measure the windings of $f$ around the non-contractible two-spheres in
$\gt$ Poincar\'e dual to the two-forms $\omega^{k}$. In (\ref{c}) it is
useful (but not mandatory) to think of $f$ as a map into the coadjoint orbit
${\cal O}^{k}\sim\gt$ through $\a^{k}$ and we will henceforth make use of this
identification. Let us therefore write $f$ as $f=h^{-1}\a^{k}h$ where $h$ is
some (not necessarily continuous) function defined by a global section of the
pull-back bundle $f^{*}(\bG)$ over $\Sigma$ (e.g.~via the local
diagonalizing functions $\{\ha\}$). Then, as a consequence of
$df = [f,h^{-1}dh]$, (\ref{c}) can be written more explicitly as
\bea
n^{k}(f) &=& \trac{1}{4\pi}\int_{\Sigma}\Tr h^{-1}\a^{k} h
 [h^{-1}dh,h^{-1}dh]\nonumber\\
         &=& \trac{1}{4\pi}\int_{\Sigma}\Tr \a^{k} [dh h^{-1},dh h^{-1}]\;\;.
         \label{gtwn}
\eea
Note that this vanishes if $h$ is globally defined as
$[dh h^{-1},dh h^{-1}] = 2d(dh h^{-1})$ is then exact, in agreement with
the fact that $f$ should have winding number zero in that case.

The expression for the winding number in the $SU(2)$-case given in (\ref{wno})
is not manifestly of the above form, so let us show that
the two definitions nevertheless agree
in that case. Learning how to reduce (\ref{wno}) to (\ref{gtwn}) will also
allow us to extend the generalized winding number (\ref{gno}) to groups
other than $SU(2)$ as verification of the homotopy invariance of (\ref{wno})
requires the use of identities which are special to $SU(2)$.
Thus, in (\ref{wno}) write $f=h^{-1}\mu h$. Using the
$SU(2)$ identity $\Tr [a,b] [c,d] = 4(\Tr ac \Tr bd - \Tr ad \Tr bc)$ one
finds
\be
\Tr f [df,df] = 4\Tr \mu^{2} \Tr \mu [dh h^{-1},dh h^{-1}]\;\;,
\ee
so that indeed for $\mu=\sigma_{3}$
the two expressions (\ref{wno}) and (\ref{gtwn}) for the winding numbers
of maps into $S^{2}\sim SU(2)/U(1)$ agree.

\subs{Generalized Winding Numbers}

On the other hand we know that torus bundles
over a surface $\Sigma$ are also classified in terms of $r$ first Chern classes
$c_{1}^{k}(A^{\lt})$, where the $A^{\lt}$ are connections on these bundles.
The formula here is
\be
c_{1}^{k}(A^{\lt}) = \trac{1}{2\pi}\int_{\Sigma} da^{k} \label{c1}
\ee
where $a^{k} = -\Tr \a^{k} A^{\lt}$ is the $k$'th `component' of $A^{\lt}$.
As the integers $ c_{1}^{k}(A^{\lt})$
determine the bundle completely, one would expect a relationship with the
integers $ n^{k}(f)$. We will now show that these two descriptions of the
torus bundles are bridged by the considerations involved in establishing
that one may conjugate maps into the torus.

To that end we introduce
a formula that interpolates between (\ref{c}) and (\ref{c1}) and
which generalises that given for the case of $SU(2)$ in (\ref{cno}).
We first put the expression for the generalized $SU(2)$ winding number
into a form which is amenable to generalization to other groups.
Writing $f$
as  $f= h^{-1}\mu h$ and using the above trace identity for $SU(2)$ and
\bea
d_{A}f&\equiv& df + [A,f] \nonumber\\
      &=& h^{-1} [A^{h^{-1}},\mu] h\;\;,
\eea
one finds that (\ref{gno},\ref{cno}) can be written as
\be
n(f,A) = -\trac{1}{2\pi}\int \Tr \mu d(A^{h^{-1}})\;\;,
\ee
(cf.~(\ref{chno})). Thus, for $A$ a gauge field on the trivial principal
$\bG$ bundle we are led to define
\be
n^{k}(f,A) = -\trac{1}{2\pi}
\int_{\Sigma} \Tr \alpha^{k} d (A^{h^{-1}}) \label{ggwn}
\ee
as the generalized winding numbers of $f\sim h^{-1}\a^{k} h$. We will see
in Corollary 5 below that they can be interpreted as monopole numbers. As
such they should provide integral representations for the magnetic numbers
introduced in \cite{gno}.

While (\ref{ggwn}) is
a very compact way of writing the winding number, there are two alternative
expressions (corresponding to (\ref{gno}) and (\ref{cno}) respectively)
which make one or the other of the properties of (\ref{ggwn}) manifest.
First of all, as in the
$SU(2)$ case, this generalized winding number differs from the winding number
$n^{k}(f)$ only by a total derivative,
\be
n^{k}(f,A) = n^{k}(f) - \trac{1}{2\pi} \int_{\Sigma}d (\Tr fA)\;\;.\label{tod}
\ee
Furthermore, it can also be written in terms of the curvature $F_{A}$ of $A$
and the covariant derivative of $f$. We write $d_{A}f$ as $d_{A}f=[D(A,f),f]$,
so that $D(A,f) = A - h^{-1}dh$ modulo terms that commute with $f$.
Then $n^{k}(f,A)$ can be written as
\be
n^{k}(f,A) = -\trac{1}{4\pi}\int \Tr f[D(A,f),D(A,f)] -\trac{1}{2\pi}
\int\Tr f F_{A}\;\;,\label{cno2}
\ee
which makes its analogy with (\ref{cno}) manifest. This also shows that
the generalized winding number makes sense for non-trivial principal
bundles. The following Proposition lists the main properties of (\ref{ggwn}).

\noindent {\bf Proposition 5:} Let $f\in\sgt$ be a smooth map and
denote by $h\in\sg$ a (possibly discontinuous) lift of $f$ to $\bG$.
Let $A\in\Omega^{1}(\Sigma,\lg)$ represent a gauge field on the trivial
principal $\bG$ bundle $P_{G}\sim \Sigma\times\bG$. Then $n^{k}(f,A)$ has
the following properties:
\begin{enumerate}
\item $n^{k}(f,A)$ is independent of $A\in\Omega^{1}(\Sigma,\lg)$
\item $n^{k}(f,A)$ is gauge invariant, i.e.
\[ n^{k}(g^{-1}f g, g^{-1} A g + g^{-1} dg ) = n^{k}(f,A) \]
for any $g\in\sg$.

\item If $h$ can be chosen to be smooth, $n^{k}(f,A)=0$.

\item For $A=0$, $n^{k}(f,A)$ reduces to the integral of the
 pull-back of the Kirilov-Kostant form $\omega^{k}$ to $\Sigma$, i.e.~to
 the winding number (\ref{c}) of $f$,
 \[n^{k}(f,A=0) = n^{k}(f)\;\;.\]
\end{enumerate}

\noindent {\bf Proof:} Properties 1 and 4 follow immediately from (\ref{tod})
and property 2 from (\ref{cno2}). Alternatively, one can argue as follows.
The variation of (\ref{ggwn}) with respect to $A$ is
\bea
\d n^{k}(f,A) &=& -\trac{1}{2\pi}\int_{\Sigma}d (\Tr \a^{k} h \d A h^{-1})
\nonumber\\
&=& - \trac{1}{2\pi} \int_{\Sigma} d (\Tr f \d A) = 0\;\;,
\eea
as $f$ and $\d A$ are globally defined. This establishes property 1. Note
that this argument also goes through for $A$ a connection on a non-trivial
$\bG$ bundle $P_{G}$.
Property 2 follows from the observation that $f\ra g^{-1} f g$
corresponds to $h\ra hg$ so that $A^{h^{-1}}$ is the manifestly gauge invariant
combination of $A$ and $h$. Property 3 holds because the integrand
of (\ref{ggwn}) is globally exact if $h$ is smooth, and property 4 is
a consequence of
\be
\Tr \a^{k} d(dh h^{-1}) = \trac{1}{2}\Tr \a^{k} [dh h^{-1}, dh h^{-1}]\;\;.
\ee
This completes the proof of the proposition an immediate consequence of which
is the equality of the winding
numbers (\ref{c}) and the torus bundle Chern classes (\ref{c1}) claimed above:

\noindent {\bf Corollary 4:} Let $f\in\sgt$, $h\in\sg$ and $A$
be as above. Define the torus gauge field $A^{\lt}$ to be the $\lt$
component of $A^{h^{-1}}$. Then the winding numbers of $f$ are equal to the
Chern numbers of $A^{\lt}$,
\[n^{k}(f) = c_{1}^{k}(A^{\lt}) \;\;.\]

Returning to our problem of conjugating maps into the torus, we can now
read off directly from the above that a smooth
map $g\in\sgr$ can be smoothly conjugated into the torus iff
the (generalized) winding number of $f$ is zero.
Furthermore, if
one insists on conjugating into the torus nevertheless, albeit by a
non-continuous $h$, the resulting map $f$ is a constant map (with winding
number zero) but $n^{k}(f,A)$ will remain unchanged, measuring the
obstruction to doing this smoothly. This establishes

\noindent{\bf Corollary 5:} Let $g\in\sgr$ be a smooth regular map,
$(f,t)$ a lift of $g$ to $\gt\times \bT_{r}$, $P_{T}$ the corresponding
$\bT$-bundle. Then the generalized winding numbers $n^{k}(f,A)$ are the Chern
numbers of $P_{T}$ and $g$ can be smoothly conjugated into $t$ iff
the $n^{k}(f,A), k = 1,\ldots,r$ are zero for some (and hence all)
$A\in\Omega^{1}(\Sigma,\lg)$.

\section{Generalizations: Non-Regular Maps and Non-Trivial $\bG$ Bundles}

In this section we will take a look at some of the topics we have only
touched briefly or glossed over completely so far. In particular, we
will extend the analysis of section 5 from regular maps to (regular)
sections of non-trivial Ad-bundles, and we will come back to the question
of non-regular maps we had quickly abandoned after the first example of
section 3. As it turns out, these two issues are not unrelated as there
may be obstructions to finding {\em any} regular sections.

\subs{Diagonalizing Sections of Non-Trivial Ad-Bundles}

We consider now the situation where the bundle $P_{G}$ is possibly non-trivial
and characterized by a set of transition functions $\{\gab\}$ with
respect to a contractible open covering $\{\ua\}$ of the base space $M$.
We furthermore assume the existence of a regular section $g = \{\ga\}$
of ${\rm Ad}P_{G}$, i.e.~a section such that all the $\ga$ take values
in $\bG_{r}$. As $\bG_{r}$ is invariant under conjugation, the notion of
a regular section is independent of the choice of local trivialization
and hence well defined. We will see below, however, that the assumption
that a regular section exists is non-trivial and imposes some topological
restrictions on $P_{G}$ (which turn out to be precisely those which permit
the regular sections to be diagonalized).

Since $g$ is a section of the adjoint bundle, its local representatives
are related on overlaps $\uab$ by
\be
\ga = \gab \gb\gab^{-1}\;\;.\label{n1}
\ee
Locally the situation is exactly as in section 5 and hence we can assume the
existence of smooth local diagonalizing functions $\ha\in\uag$ such that
$\ha\ga\ha^{-1} =\ta$ takes values in $\bT_{r}$ (this has already been
established in Proposition 1). It then follows from
(\ref{n1})  that
\be
\ha^{-1}\ta\ha = \gab \hb^{-1}\tb\hb\gab^{-1}\;\;,
\ee
or, that on overlaps the $\ta$ are related by
\be
\ta = (\ha\gab\hb^{-1}) \tb (\ha\gab\hb^{-1})^{-1}\;\;.\label{n2}
\ee
We can now argue exactly as in section 5 to conclude that, as the $\ta$
are regular,
the (transition) functions $\ha\gab\hb^{-1}$ take values in $N(\bT)$.
Moreover, if $\bG$ is simply connected one can use the ambiguity
$\ha\ra n_{\a}\ha$ with $n_{\a}:\ua\ra
N(\bT)$ to conjugate all the $\ta$ into the same fundamental domain
$D\sim \bT_{r}/W$. We can then conclude from (\ref{n2}) that
the $\ha\gab\hb^{-1}$ can actually be chosen to take values in $\bT$,
\be
\ha\gab\hb^{-1}:\uab\ra \bT\;\;,\label{n3}
\ee
and that the locally defined diagonalized maps $\ta$ piece together to
a globally well defined $\bT_{r}$-valued function $t=\{\ta\}$,
\be
\ta = \tb \;\;\;\;{\rm on}\;\;\;\;\uab\;\;.\label{n4}
\ee
These results are the precise counterparts of (\ref{tr1}) and
(\ref{tr2}) obtained
in section 5 in the case of trivial bundles where we interpreted them
in terms of restrictions of the structure group of a trivial principal
$\bG$ bundle. Here we appear to reach the conclusion that any $\bG$ bundle
can be restricted to a $\bT$ bundle which can obviously not be correct.
We will come back to this below.

In analogy with (\ref{tr4}) we can also define local $\gt$-valued maps
$f_{\a}= h_{\a}^{-1}\mu h_{\a}$. However, unlike in the case considered
there, these do not automatically piece together to give a globally well
defined map into $\gt$. Rather, on overlaps, they transform as
\be
f_{\a} = \gab f_{\b} \gab^{-1}\;\;,\label{n5}
\ee
This means that the $\{f_{\a}\}$ define a global section of the
homogenous bundle $E_{G/T}$ with fibre $\gt$ to be introduced below.

As one expects $t$ to define a section of the adjoint bundle of some
principal $\bT$ bundle $P_{T}$ which is trivial for any $P_{T}$
(the adjoint action of $\bT$ on itself is trivial), the result
(\ref{n4}) is eminently reasonable. The crux of the matter lies in
the conclusion (\ref{n3}) which cannot be fullfilled in general and
which implies that some non-innocuous topological assumption has slipped
into our above analysis. Alternatively, the possible non-existence of
global sections of $E_{G/T}$ constitutes a further obstruction to smoothly
and globally conjugating the regular section $g = \{\ga\}$ into $\bT$.
To better understand this point we make a brief
excursion concerning the restriction of structure groups.

\subs{Regular Sections and Restrictions of the Structure Group}

Let $P_{G}$ be a principal $\bG$ bundle and $\bH$ a subgroup of
$\bG$. One says that the structure group of $P_{G}$ can be restricted to
$\bH$ if there exists a principal $\bH$ bundle $P_{H}$ and an embedding
$j:P_{H}\hookrightarrow P_{G}$ whch is a strong (i.e.~fiber preserving)
principal bundle morphism which induces the embedding $i:\bH\hookrightarrow\bG$
on the fibers. Alternatively, in terms of local coordinates and transition
functions $\gab$ of $P_{G}$, $P_{G}$ is said to have a restriction to $\bH$
if there exist functions $\ha\in\uag$ such that the equivalent
transition functions $\ha\gab\hb^{-1}$ (corresponding to a change of local
trivialization) take values in $\bH$.

Two of the more elementary results concerning restrictions of structure
groups are that the structure group can always be restricted to a
maximal compact subgroup and that it can be restricted to the trivial
group $\{1\}$ (and hence any subgroup of $\bG$)
iff $P_{G}$ is trivial. The latter already shows that
in general there may be (topological) obstructions to the restriction of
structure groups.

To describe the general situation when
$\bH$ is a non-trivial subgroup of $\bG$ we need to introduce the quotient
space $E_{G/H} = P_{G}/\bH$ which is well defined since $\bG$ (and hence $\bH$)
acts freely on the right on $P_{G}$. $E_{G/H}$ is a fiber bundle over $M$
with typical fiber $\gt$ associated to $P_{G}$ via the action of $\bG$ on
$\gt$. Then the fundamental result concerning restrictions of structure
groups is \cite{hu} that there is a bijective correspondence between

- principal $\bH$ bundles $P_{H}$ which are restrictions of $P_{G}$, and

- global sections of the associated bundle $E_{G/H}$.

\noindent As the proof of this result is not too hard and gives some
insight into the manipulations performed in section 5 and above, we give
a sketch of it here. First of all, given a section $s:M\ra E_{G/H}$, one
can use it to pull back the principal $\bH$ bundle $P_{G}\ra E_{G/H}$
to $M$. The resulting principal $\bH$ bundle $P_{H}$ over $M$ is easily
seen to satisfy all the requirements of a restriction. Conversely,
given a restriction $(P_{H},j)$, one defines a section of $E_{G/H}$ by
composing $j:P_{H}\hookrightarrow P_{G}$ with the restriction of the
projection map
\be
\chi:P_{G}\times \gt \rightarrow P_{G}\times_{\bG}\gt = E_{G/T}
\ee
to the origin $o\in\gt$,
\be
s(x):= \chi(j(p),o)\;\;.
\ee
Here $p$ is any point in the fiber of $P_{H}$ above $x\in M$ and the
right hand side does not depend on the choice of $p$ because $j$ is
by assumption a principal bundle morphism, $j(ph) = j(p)i(h)$ for $h\in\bH$.

Let us now compare this with what we did in sections 4 and 5 under the
assumption that $P_{G}$ is trivial. Then $E_{G/T}$ is trivial as well and
therefore there are no obstructions to restricting the structure group of
$P_{G}$ to $\bT$. Restrictions simply correspond to maps from $M$ to $\gt$,
homotopic maps giving rise to isomorphic principal $\bT$ bundles over $M$.
This is just what we found in a more pedestrian way in section 5; see in
particular equation (\ref{tr2}) which shows that the structure group of
$P_{G}\sim M\times\bG$ has been reduced to $\bT$, and (\ref{tr3}) which
exhibits the corresponding map into $\gt$. Furthermore, if there are
no non-trivial $\bG$ bundles on $M$, every principal $\bT$ bundle $P_{T}$
arises as the restriction of $P_{G}$ for some map $f:M\ra\gt$, as the
transition functions $\tab$ of $P_{T}$ can {\em a fortiori} be regarded as
the transition functions of a $\bG$ bundle by composing them with $i:
\bT\hookrightarrow\bG$.

In general, however, when $P_{G}$ and $E_{G/T}$ are non-trivial, there will
be topological obstructions to the existence of global sections of $E_{G/T}$
and hence to restrictions of the structure group, the primary obstructions
typically lying in $H^{k}(M,\pi_{k-1}(\gt))$ for some $k$. What
(\ref{n3}) shows, on the other hand, is that a restriction to $\bT$
exists if ${\rm Ad}P_{G}$ has a regular section while (\ref{n5}) exhibits the
corresponding section of $E_{G/T}$. As the converse, if the bundle admits
a restriction then there is a regular section, is easily established in
general (by following the reasoning leading to (\ref{conv1}) and
(\ref{conv2})), we can summarize the consequences of the above considerations
in

\noindent {\bf Proposition 6:} Let $P_{G}$ be a principal $\bG$ bundle
over $M$ and $E_{G/T}=P_{G}/\bT$ its associated homogeneous bundle.
$P_{G}$ admits a restriction to $\bT$ (equivalently, $E_{G/T}$ has
a global section) if and only if ${\rm Ad}P_{G}$ has a regular section.

In a sense this is the central result of this paper. It explains the intimate
relationship we found between  diagonalization and
restriction of the structure group and it highlights the crucial role played
by the assumption of regularity.

Nevertheless this result may seem to be somewhat curious as {\em a priori} the
condition of regularity is not a cohomological condition while it nevertheless
implies that there are no topological obstructions to the existence of a
global section of $E_{G/T}$. However, it is not unlike the relation between
the triviality of a line bundle and the existence of a nowhere vanishing
section in that an algebraic condition has a topological implication.

It would be nice to have a demonstration
of Proposition 6 which does not rely on diagonalization but deals directly
with the obstructions instead, but we have been unable to find such a direct
proof. In four dimensions, however, necessary and sufficient conditions
for the existence of restrictions of $SU(n)$ bundles can be read off more or
less by inspection and this gives some insight into the nature of this
problem.

We recall first that $SU(n)$ bundles $P$ on a compact oriented
four-manifold are
completely classified by the second Chern class $c_{2}(P)\in H^{4}(M,\ZZ)
\sim \ZZ$. In terms of the curvature $F_{A}$ of a connection $A$ on $P$
the Chern-Weil representative of $c_{2}(P)$ is
\be
c_{2}(P) = \trac{1}{8\pi^{2}}\int_{M}\Tr F_{A}F_{A}  \label{chern2}
\ee
(with the trace normalized to $\Tr\lambda^{a}\lambda^{b}=2\d^{ab}$, the
$\lambda^{a}$ a basis of the Lie algebra of $SU(n)$). Torus bundles $P_{T}$,
$\bT\sim U(1)^{n-1}$, on the other hand are classified by
$H^{2}(M,\ZZ^{n-1})$. As all $\bT$ bundles can be regarded as $SU(n)$ bundles,
they will all arise as the restriction of some $SU(n)$ bundle but not
necessarily as restrictions of the trivial $SU(n)$ bundle. Moreover,
some $SU(n)$ bundles may have no restrictions at all while others may admit
several inequivalent restrictions. In this four-dimensional context it
is straightforward to find obstructions to such an Abelianization.
Let us first write $\Tr F_{A}F_{A}$ locally as
\be
\Tr F_{A}F_{A} = d \Tr (AdA + \trac{2}{3}A^{3})\;\;.\label{cs}
\ee
If one has been able to abelianize (with transition functions as in
(\ref{n3})), then one may as well write
\be
\Tr F_{A} F_{A} =
   d \Tr A^{\lt} d A^{\lt} + d \Tr A^{\lk} d_{A^{\lt}} A^{\lk}\;\;.\label{cs2}
\ee
As the second term transforms homogeneously under gauge transformations
(see (\ref{connt}))
and hence under change of local trivializations, the second term is globally
defined and does not contribute to the integral (\ref{chern2}). Hence one
finds that for a principal $SU(n)$ bundle which admits a restricition to a
$\bT$ bundle $P_{T}$, its second Chern class is related to the curvature
of a connection on $P_{T}$ by
\be
c_{2}(P) = \trac{1}{8\pi^{2}}\int_{M}\Tr dA^{\lt} dA^{\lt}\;\;.\label{rel1}
\ee
By looking at some concrete examples of four-manifolds we will see that
this relation can impose severe constraints on $c_{2}(P)$.

Let us, for instance, take $M$ to be the four-sphere $M=S^{4}$. Then there
are no non-trivial $\bT$ bundles on $M$ as $H^{2}(M,\ZZ)=0$,
and the right hand side of (\ref{rel1})
is zero as the integrand is then necessarily globally exact. Hence we reach
the conclusion that only the trivial $SU(n)$ bundle on $S^{4}$ admits a
restriction to a $\bT$ bundle (the trivial $\bT$ bundle in this case). This
may also be seen in a differnet way by noting that, on any $n$-sphere, the
bundle is characterized by the glueing (transition) function $h$ from the
equator $\sim S^{n-1}$ to the group $\bG$. If $h$ takes values in $\bT$,
then its winding number is zero ($\pi_{n-1}(\bT) = 0$ for $n>2$) and hence
\be
8 \pi^{2} c_{2}(P) = \int_{S^{3}}\Tr (h^{-1}d h)^{3} = 0\;\;.
\ee
Thus we conclude that the adjoint bundles of non-trivial $SU(n)$ bundles
over $S^{4}$ have no regular sections whatsoever.

This is not to mean that only trivial $SU(n)$ bundles can be reduced to
$\bT$ bundles. As another example consider $M=\CC\PP^{2}$ and $\bG=SU(2)$.
In this case, $H^{2}(M,\ZZ)\sim H^{4}(M,\ZZ) \sim \ZZ$,
generated by the K\"ahler form $\omega$.
Thus there are non-trivial torus and $SU(2)$ bundles on $\CC\PP^{2}$.
The curvature of the connection on a $U(1)$ bundle is cohomologous to
$k\omega$  for $k\in \ZZ$ and, as $\omega^{2}[\CC\PP^{2}]=1$, a necessary
condition for an $SU(2)$ bundle $P$ to be reducible to $U(1)$ is that
$c_{2}(P)=k^{2}$ for some $k\in \ZZ$. As any $U(1)$ bundle with first
Chern class $k$ is the reduction of some $SU(2)$ bundle, this condition is
also sufficient and for every non-trivial $SU(2)$ bundle on $\CC\PP^{2}$
there are two inequivalent reductions to $U(1)$, characterized by the
first Chern class $\pm k$.

This situation is more or less the same for all compact four-manifolds.
If a torus bundle, thought of as an $SU(n)$ bundle, has second Chern class
$c_{2}=m$, then it can be obtained as the reduction of this $SU(n)$ bundle.
Conversely, if an integer $m$ does not arise as the second Chern class of
some torus bundle, the corresponding $SU(n)$ bundle with $c_{2}(P)=m$
cannot be Abelianized. As a consequence of Proposition 6 such bundles
have no regular sections whatsoever.

By the above reasoning one can establish in general that if a principal
$\bG$ bundle $P_{G}$ has a restriction to a principal $\bf H$ bundle $P_{H}$,
where $\bf H$ is any subgroup of $\bG$ containing $\bT$, then the
characteristic
classes of these bundles will be the same. While this is more or less obvious
on general grounds, the considerations involving diagonalization (or
conjugation into $\bf H$ in the more general case) permit one to be quite
explicit about this.

There is one further complication that arises when $M$ admits non-trivial
$\bG$ bundles, already implicit in the above discussion.
It is certainly still true that every
$\bT$ bundle on $M$ will arise as the restriction of some $\bG$ bundle.
However, a given principal $\bG$ bundle $P_{G}$ will only give rise to those
principal $\bT$ bundles after diagonalization of its regular sections which
arise as restrictions of $P_{G}$. This will have to be reflected in the
corresponding Weyl integral formula which will then include a sum over
only a restricted class of isomorphism classes of principal $\bT$ bundles
(unless, of course, the original theory is defined as a sum over all
isomorphism classes of $\bG$ bundles).

\subs{Are Regular Maps Generic?}

While we have seen above that non-trivial adjoint bundles may admit
no regular sections at all, which forces us face the task of diagonalizing
non-regular maps, one may have hoped that at least for trivial bundles
regular maps are generic so that `most' maps can indeed be conjugated
into smooth torus-valued functions by the results of sections 4 and 5,
at least via locally defined or discontinous diagonalizing functions $h$.
This turns out to be so for Lie algebra valued maps but a simple
example will show that it is not necessarily true for group valued
maps.

Let us look at the Lie algebra case first. If $P_{G}$ is trivial,
sections of the adjoint bundle ${\rm ad}P_{G} = P_{G}\times_{\rm ad}\lg$,
a vector bundle over $M$ with fiber $\lg$, can be identified with maps
from $M$ to $\lg$. Now the non-regular points in $\lg$ form a set of
codimension at least three: the non-regular elements of $\lt$ form a set
of codimension one (dimension $r-1$), as they partition $\lt$ into its
Weyl alcoves; the dimension of a coadjoint orbit through a non-regular element
is strictly smaller than the dimension $\dim\bG -r$ of $\gt$ and in fact
at most $\dim\bG - r -2$ because the orbit is symplectic and hence even
dimensional; hence the dimension of the set of non-regular elements is at
most
\be
\dim (\lg\setminus\lg_{r})\leq (r-1) + (\dim\bG -r -2) = \dim\bG - 3\;\;.
\ee
It follows that regular maps are indeed generic in any dimension.

Because of topological complications not present in the Lie algebra case,
the corresponding statement for group valued maps is false. To see that,
let us consider as a simple example the space of maps from $M=S^{3}$ to
$\bG = SU(2)\sim S^{3}$. This space consists of an infinite number of
connected components labelled by the winding number of the map in
$\pi_{3}(SU(2))=\ZZ$. As the only non-regular elements of $SU(2)$ are
plus or minus the identity, regular maps are those which avoid the
north and south poles of the target $S^{3}$. Clearly generic maps in
the zero winding number sector have this property. As there
are no non-trivial $U(1)$ bundles on $S^{3}$, $H^{2}(S^{3},\ZZ)=0$,
any such map can be globally and smoothly conjugated into $U(1)$.
On the other hand, any map in one
of the other sectors has in particular the property that its image is the
entire $SU(2)$, covered an appropriate number of times. Hence, no map with
a non-trivial winding number can be regular.

The upshot of this is that neither regular $\bG$-valued maps
nor regular sections of non-trivial ${\rm Ad}P_{G}$ or ${\rm ad}P_{G}$
bundles can be expected to be generic in general, the only exception
being $\lg$-valued maps.

\subs{Diagonalization of Non-Regular Maps?}

The fact that even for trivial bundles there may be too many
non-regular maps for comfort provides an additional impetus for
coming to terms with the diagonalization of these maps. Unfortunately,
this problem appears to be much harder than the corresponding one
for regular maps and in the following we will only make a few remarks
and tentative suggestions in that direction.

Let us first recall at which points in our analysis the assumption of
regularity entered (we take $\bG$ to be simply connected):

\begin{enumerate}
\item The fact that the conjugation map $q:\gt\times\bT\ra\bG$, (\ref{w1}),
      is proper and, in fact, a (trivial) fibration away from
      the non-regular points allowed us to solve the first lifting problem
      in section 4 for regular maps.
\item Regularity of $g$ (and hence of the $\ta$) allowed us to conclude from
      (\ref{overlap}) that the transition functions $\hab=\ha\hb^{-1}$
      take values in $N(\bT)$ (and can even be chosen to reduce the
      structure group to $\bT$).
\end{enumerate}

If $g\in\mg$ is not regular then clearly no such restriction will necessarily
be imposed on the transition functions defined by the local diagonalizing maps.
E.g.~if $g$ is the constant identity map, the $\ha$ are completely arbitrary.
As this $g$ is already diagonal, this may not be too much of a concern, but
other problems arise for maps which take on both regular and non-regular
values as we have already seen in example 1 of section 3. For instance, the
triviality of the fibration (\ref{w1}) may alternatively be expressed by saying
that the quotient of $\bG_{r}$ by the adjoint action of $\bG$ is a smooth
manifold,
\be
\bG_{r}/{\rm Ad}\bG \sim \bT_{r}/W
\ee
and that topologically (and smoothly) one has
\be
\bG_{r}\sim \bT_{r}/W \times\gt\;\;,
\ee
the points in $D\sim \bT_{r}/W$ labelling the regular (maximal) coadjoint
orbits in $\bG_{r}$. If one considers maps taking values in all of $\bG$,
one has to come to terms with the fact that the quotient
\be
\bG/{\rm Ad}\bG\sim \bT/W
\ee
is not a smooth manifold (but the closure of a Weyl alcove or, rather, its
image under the exponential map), and that
the fiber of $\bG\ra\bG/{\rm Ad}\bG$ above a singular (non-regular) point is
isomorphic to the coadjoint orbit through that non-regular point and hence
strictly
smaller than that at a regular point. Clearly this is a rather singular
situation to consider and accounts for most of the problems associated
with non-regular maps.

Looking back at example 1 of section 3 we see that the failure to be smoothly
diagonalizable is due to the combined effect of having a non-regular map and
a non-simply connected base space, the diagonalized map $t(x)$ being well
defined on the non-trivial double cover of the circle. This and the fact
that there are no
non-trivial $W$ bundles on simply connected manifolds suggest that it may be
be possible to prove stronger statements regarding non-regular maps in
that case.

\section{Applications: A Weyl Integral Formula for Path Integrals}

In the previous sections we have analyzed the problem of diagonalizing
maps from a manifold $M$ into a compact Lie group $\bG$ or its Lie algebra
$\lg$. As mentioned in the Introduction, this problem arose in a field
theoretic context when we attempted to exploit the rather large local
gauge symmetry present in certain low-dimensional non-Abelian gauge theories
to abelianize (and hence more or less trivialize) the theories via
diagonalization \cite{btver,btlec,ew}. Assuming that the contributions from
non-regular maps can indeed be neglected in these examples (and we have
nothing to add to the arguments put forward in \cite{btlec} to that effect),
the analysis of the present paper can be regarded as a topological
justification
for the formal path integral version of the Weyl integral formula we used
to solve these theories.

The Weyl integral formula expresses the integral of a smooth
(real or complex valued) function over $\bG$ in terms of an integral over
$\bT$ and $\gt$, using the conjugation map $q$ (\ref{w1}) to pull back
the Haar measure on $\bG$ to $\gt\times\bT$ and reads
\be
\int_{\bG}\! dg\,f(g) = \int_{\bT}\! dt\,\Delta(t)
\int_{\gt}\! dg\,f(g^{-1}tg)\label{weyl1}\;\;.
\ee
Here $\Delta(t)$, the Weyl determinant, is the Jacobian of $q$. Its precise
form will not interest us here and we just note that it vanishes precisely
at the non-regular points of $\bT$ (this being the mechanism by which
contributions form non-regular points should be suppressed in the functional
integral). For an explanation of the standard proof
of (\ref{weyl1}) and for a derivation in the spirit of the Faddeev-Popov
trick see \cite{btver,btlec}.

The case of interest to us is when the function $f$ is conjugation invariant
(a class function), i.e.~when $f$ satisfies
\be
f(h^{-1}gh) = f(g) \;\;\;\;\;\;\forall\;\;g,h\in\bG\;\;.
\ee
In that case, since any element of $\bG$ is conjugate to some element of
$\bT$, both $f$ and its integral over $\bG$ are determined by their
restriction to $\bT$ and the Weyl integral formula reflects this fact,
\be
\int_{\bG}\! dg\,f(g) = \int_{\bT}\! dt\,\Delta(t)f(t)
\label{weyl2}\;\;.
\ee
It is this formula which we would like to generalize to functional integrals,
i.e.~to a formula which relates an integral over a space of maps into $\bG$
to an integral over a space of maps into $\bT$.

For concreteness, consider a local functional $S[g;A]$ (the `action')
of maps $g\in\mg$ and gauge fields $A\in\Omega^{1}(M,\lg)$, i.e.~of
sections of ${\rm Ad}P_{G}$ and connections on a trivial principal $\bG$
bundle $P_{G}\sim M\times \bG$ (a dependence on other fields could be
included as well). Assume that $\exp iS[g;A]$ is gauge invariant,
\be
\exp i S[g;A] = \exp i S[h^{-1}gh; A^{h}] \;\;\;\;\;\;\forall\;h\in\mg\;\;,
\label{weyl3}
\ee
at least for smooth $h$. If e.g.~a partial integration is
involved in establishing the gauge invariance (as in Chern-Simons theory),
this may fail for
non-smooth $h$'s and more care has to be exercised when such a gauge
transformation is performed.
Then the functional $F[g]$ obtained by integrating  $\exp iS[g;A]$ over $A$,
\be
F[g]:=\int\!D[A]\,\exp iS[g;A]\;\;,
\ee
is conjugation invariant,
\be
F[h^{-1}gh] = F[g]\;\;.
\ee
It is then tempting to use a formal analogue of (\ref{weyl2}) to reduce the
remaining integral over $g$ to an integral over maps taking values in the
Abelian group $\bT$. In field theory language this amounts to using the
gauge invariance (\ref{weyl3}) to impose the `gauge condition' $g(x)\in\bT$.
The first modification of (\ref{weyl2}) will then be the replacement of
the Weyl determinant $\Delta(t)$ by a functional determinant $\Delta[t]$
of the same
form  which needs to be regularized appropriately (see the Appendix of
\cite{btlec}).

However, the main point of this paper is that this is of course not the
whole story. We already know that this `gauge condition' cannot necessarily
be achieved smoothly and globally. Insisting on achieving this `gauge'
nevertheless, albeit via non-continuous field
transformations, turns the $\lt$-component $A^{\lt}$
of the transformed gauge field $A^{h^{-1}}$ into a gauge field on a
possibly non-trivial $\bT$ bundle $P_{T}$ (while the $\lk$-components
transform as sections of an associated bundle). Moreover we know that
all those $\bT$ bundles will contribute which arise as restrictions of
the (trivial) bundle $P_{G}$. Let us denote the set of isomorphism classes
of these $\bT$ bundles by $[P_{T};P_{G}]$. Hence the `correct'
(meaning correct modulo the analytical difficulties inherent in making any
field theory functional integral rigorous) version of the Weyl integral
formula, capturing the topological aspects of the situation,
is one which includes a sum over the contributions from the
connections on all the isomorphism classes of bundles in $[P_{T};P_{G}]$.

Let us denote
the space of connections on $P_{G}$ and on a principal $\bT$ bundle $P_{T}^{l}$
representing an element $l\in [P_{T};P_{G}]$ by ${\cal A}$ and ${\cal A}[l]$
respectively and the space of one-forms with values in the sections of
$P_{T}^{l}\times_{\bT}\lk$ by ${\cal B}[l]$.
Then, with
\be
Z[P_{G}] = \int_{\cal A}\! D[A]\int\! D[g]\exp i S[g;A] \;\;,
\ee
the Weyl integral formula for functional integrals reads
\be
Z[P_{G}] = \sum_{l\in [P_{T};P_{G}]}\int_{{\cal A}[l]}\! D[A^{\lt}]
                         \int_{{\cal B}[l]}\! D[A^{\lk}]
                         \int\! D[t] \Delta[t] \exp i S[t;A^{\lt},A^{\lk}]
                         \label{weyl4}
\ee
(modulo a normalization constant on the right hand side).
The $t$-integrals carry no $l$-label as the spaces
of sections of ${\rm Ad}P_{T}^{l}$ are all isomorphic to the space
of maps into $\bT$.

In the examples considered in \cite{btver,btlec}, Chern-Simons theory on
three-manifolds of the form $\Sigma\times S^{1}$, 2d Yang-Mills theory
and the $\bG/\bG$ gauged Wess-Zumino-Witten model, the fields $A^{\lk}$
entered purely quadratically in the reduced action $S[t;A^{\lt},A^{k}]$
and could be integrated out directly, leaving behind an effective Abelian
theory depending on the fields $t$ and $A^{\lt}$ with a measure determined
by $\Delta[t]$ and the (inverse) functional determinant coming from the
$A^{\lk}$-integration. The general structure of these terms and the
`quantum corrections' coming from the regularization has been determined in
\cite{ew}.

A further property these models were found to have is that they localize
onto reducible connections and their isotropy groups (in the case of the
$\bG/\bG$ model) respectively algebras (for Yang-Mills theory) so that,
in practice, the necessity only ever arose to diagonalize these maps.
This is possible globally even if the group is not simply connected (when, as
we recall from section 4, the existence of a globally smooth diagonalized map
$t$ or $\tau$ is not guaranteed {\em a priori}). The reason for this is
the following (for group valued maps - the Lie algebra case is entirely
analogous):

The reducibility condition $A^{g}=A$ implies that $\Tr g^{n}$ is constant for
all $n$. This allows one to determine that $g$ is conjugate to a $t$ which
is constant globally and (of course) unique up to an overall
$W$-transformation. This provides the $\bT_{r}$ part of the lift in diagram
(4.2).  Furthermore, the constancy of the traces implies that $g$
can itself be regarded as a map into $\gt$ and hence furnishes the
$\gt$-part $f$ of the lift. At this point the argument can then proceed as
in the simply-connected case. The fact that isotropy groups of connections are
indeed conjugate to subgroups of $\bG$ (thought of as spaces of constant maps)
is well known. What seems to be less generally appreciated is the fact that
the conjugation itself cannot necessarily be done globally.

We have also applied this formula to several other models
like $BF$ theories in three dimensions (related to $3d$ gravity) and
the supersymmetric Chern-Simons models of Rozansky and Saleur \cite{rs}.
The formula
can also be used to go some way towards evaluating the generating functional
for Donaldson theory on K\"ahler manifolds with the action as in \cite{gtlec}.
These results will be presented elsewhere.

\subsubsection*{Acknowledgements}
We are grateful to E.~Witten for alerting us to the problem addressed in
this paper, and to M.S.~Narasimhan for his interest and a careful reading of
the manuscript. We also thank A.~d'Adda for discussions.

\rnc{\Large}{\normalsize}


\begin{thebibliography}{00}
\addcontentsline{toc}{section}{References}
\frenchspacing
\small
\addtolength{\itemsep}{-4pt}
\bibitem{btver} M. Blau and G. Thompson, {\em Derivation of the Verlinde
                Formula from Chern-Simons Theory and the $G/G$ model},
                Nucl. Phys. B408 (1993) 345-390.
\bibitem{btlec} M. Blau and G. Thompson, {\em Lectures on 2d Gauge Theories:
                Topological Aspects and Path Integral Techniques}, ICTP
                preprint IC/93/356, 70 p., available as hep-th/9310144.
                To appear in the Proceedings of the 1993 Trieste Summer
                School on High Energy Physics and Cosmology.
\bibitem{btloc} M. Blau and G. Thompson, {\em in preparation}.
\bibitem{btd}   T. Br\"ocker and T. tom Dieck, {\em Representations of
                Compact Lie Groups}, Springer, New York (1985).
\bibitem{gno}   P. Goddard, J. Nuyts and D. Olive, {\em Gauge Theories and
                Magnetic Charge}, Nucl. Phys. B125 (1977) 1-28.
\bibitem{gp}    K. Grove and G. K. Pedersen, {\em Diagonalizing Matrices
                over $C(X)$}, J. Funct. Anal. 59 (1984) 65-89.
\bibitem{hel}   S. Helgason, {\em Differential Geometry, Lie Groups, and
                Symmetric Spaces}, Academic Press, Orlando (Fl.) (1978).
\bibitem{tho}   G. 't Hooft, {\em The Topology of the Gauge Condition and New
                Confinement Phases in Non-Abelian Gauge Theories},
                Nucl. Phys. B190 (1981) 455-478.
\bibitem{hu}    D. Husemoller, {\em Fibre Bundles} (2nd Edition), Springer,
                New York (1975).
\bibitem{ps}    A. Pressley and G. Segal, {\em Loop Groups}, Oxford University
                Press, Oxford (1986).
\bibitem{rs}    L. Rozansky and H. Saleur, {\em Reidemeister Torsion, the
                Alexander Polynomial and the $U(1,1)$ Chern-Simons Theory},
                Yale preprint YCTP-P35-1992, available as hep-th/9209073.
\bibitem{gtlec} G. Thompson, {\em Topological Gauge Theory and Yang-Mills
                Theory}, in the Proceedings of the 1992 Trieste
                Summer School on {\em High Energy Physics and Cosmology}
                (eds. E. Gava et al.), World Scientific, Singapore (1993),
                1-75.
\bibitem{ew}    E. Witten, {\em The Verlinde Algebra and the Cohomology of
                the Grassmannian}, IAS preprint IASSNS-HEP-93/41, 78 p.,
                available as hep-th/9312104.

\end{thebibliography}
\end{document}